\RequirePackage{fix-cm}
\RequirePackage{rotating}
\documentclass[smallextended]{svjour3}       
\smartqed  

\def\jelname{{\bfseries JEL classification}}
\def\JEL#1{\par\addvspace\medskipamount{\rightskip=0pt plus1cm
		\def\and{\ifhmode\unskip\nobreak\fi\ $\cdot$
		}\noindent\jelname\enspace\ignorespaces#1\par}}

\usepackage[utf8]{inputenc}
\usepackage[english]{babel}
\usepackage{natbib} 
\setcitestyle{authoryear,open={(},close={)},aysep={}}

\usepackage{graphicx} 
\usepackage{epstopdf} 
\usepackage{subfigure} 
\usepackage{lscape}

\usepackage[displaymath, mathlines,running]{lineno}
\setpagewiselinenumbers 

\usepackage{booktabs}
\usepackage{tabularx}
\usepackage[flushleft]{threeparttable}
\usepackage[
singlelinecheck=false 
]{caption}
\usepackage{xcolor}
\usepackage{colortbl}
\usepackage{dcolumn} 
\newcolumntype{d}[1]{D{.}{.}{#1}}  
\newcommand\mc[1]{\multicolumn{1}{c}{#1}} 

\usepackage[colorlinks=true,linkcolor=black,anchorcolor=black,citecolor=black,filecolor=black,menucolor=black,runcolor=black,urlcolor=black]{hyperref}
\usepackage{url} 

\usepackage{amssymb}
\usepackage{bbm}
\usepackage{amsmath}

\DeclareMathOperator*{\argmin}{arg\,min}

\journalname{Financial Markets and Portfolio Management}

\begin{document}

\title{Cross-validated covariance estimators for \\ high-dimensional minimum-variance portfolios}

\titlerunning{Cross-validated covariance estimators for high-dimensional MVP}        

\author{Sven Husmann \and Antoniya Shivarova$^{\ast}$\thanks{$^\ast$Corresponding author.
		Email: shivarova@europa-uni.de} \and Rick Steinert}

\institute{Sven Husmann \and Antoniya Shivarova \and Rick Steinert  \at Europa-Universit\"at Viadrina \\ 
			Gro\ss e Scharrnstra\ss e 59 \\ 15230 Frankfurt (Oder), Germany \\ 
			}

\date{Received: date / Accepted: date}

\maketitle

\begin{abstract}
The global minimum-variance portfolio is a typical choice for investors because of its simplicity and broad applicability. Although it requires only one input, namely the covariance matrix of asset returns, estimating the optimal solution remains a challenge. In the presence of high-dimensionality in the data, the sample covariance estimator becomes ill-conditioned and leads to suboptimal portfolios out-of-sample. To address this issue, we review recently proposed efficient estimation methods for the covariance matrix and extend the literature by suggesting a multi-fold cross-validation technique for selecting the necessary tuning parameters within each method. Conducting an extensive empirical analysis with four datasets based on the S\&P 500, we show that the data-driven choice of specific tuning parameters with the proposed cross-validation improves the out-of-sample performance of the global minimum-variance portfolio. In addition, we identify estimators that are strongly influenced by the choice of the tuning parameter and detect a clear relationship between the selection criterion within the cross-validation and the evaluated performance measure.
\JEL{ G10 \and G11 \and C13 \and C80}
\end{abstract}

\section{Introduction}
\label{intro}
Based on the simple but essential idea of diversification and optimal risk-return profile of an investment strategy, the mean-variance model by \citet{Markowitz.1952} still represents the groundwork for portfolio optimization. In its original design, Markowitz portfolio theory assumes perfect knowledge about the expected value and variance of returns. For practical implementations, however, these parameters have to be estimated from historical data. The misspecifications due to error in estimation can lead to strong deviations from optimality and therefore an inferior out-of-sample performance \citep{Jobson.1981, Frost.1986, Michaud.1989, Broadie.1993}. This major drawback has been tackled from different perspectives in the financial literature. Some focus on estimation errors in the portfolio weights directly \citep[see, e.g.,][]{Brodie.2009, DeMiguel.2009b}, whereas others work on the inputs by improving expected returns and the covariance matrix.

In particular, portfolio weights are extremely sensitive to changes in expected returns \citep{Best.1991b, Best.1991}, which in turn are more difficult to estimate than the covariances of returns \citep{Merton.1980}. It is therefore not surprising that a considerable part of recent academic research focuses on the global minimum-variance portfolio (GMV), as this does not depend on expected returns.\footnote{\citet{DeMiguel.2009} additionally show that the mean-variance portfolio is outperformed out-of-sample by the minimum-variance portfolio not only in terms of risk, but as well in respect to the return-risk ratio.} However, even if investors decide to use the global minimum-variance portfolio, the estimation errors associated with the covariances can still lead to significant estimation errors in the portfolio weights, especially in a high-dimensional scenario.

We cover several approaches that have been shown to overcome these estimation issues and perform well in terms of out-of-sample variance. For instance, we discuss the linear shrinkage estimators of \citet{Ledoit.2004, Ledoit.2004b} designed to offer an optimal bias-variance trade-off between the sample covariance matrix and a structured target matrix. Furthermore, we adopt the recent nonlinear shrinkage technique by \citet{Ledoit.2017} that is proven to be optimal under a variety of financially relevant loss functions \citep{Ledoit.2018b}. Since an underlying factor structure is considered a valid assumption for asset returns' covariance estimators, we outline the elaborate principal orthogonal complement thresholding (POET) estimator by \citet{Fan.2013}. In addition, we follow the findings of the recent empirical studies by \citet{Goto.2015} and \citet{Torri.2019} and include a sparse precision matrix estimator, namely the graphical least absolute shrinkage and selection operator (GLASSO), which has so far received little attention in portfolio optimization research. To the best of our knowledge, we are the first to compare state-of-the-art covariance estimators such as the nonlinear shrinkage and POET with the GLASSO and to prove its significant outperformance in a high-dimensional scenario.

More importantly, the selected covariance estimation methods share one thing in common: a regularization of the sample covariance is performed to optimize its out-of-sample performance. For example, linear shrinkage methods need an optimal shrinkage intensity to balance the included variance and bias, whereas the performance of the GLASSO depends on the level of sparsity, induced by a penalty parameter. The procedure for optimally identifying those tuning parameters often includes the choice of a specific loss function to be minimized. As often advocated, a loss function or measure of fit in the model estimation is best aligned with the evaluation framework \citep{Christoffersen.2004, Ledoit.2017, Engle.2019}. To exploit those effects in more detail, we apply a nonparametric cross-validation (CV) technique with different selection criteria to determine the optimal parameters, necessary for the calculation of all the considered covariance estimators. 

Since we focus on enhancing the risk profile of the GMV portfolios, we choose two relevant risk-related measures for our data-driven estimation methodology and the corresponding out-of-sample performance evaluation, namely the mean squared forecasting error (MSFE), as in \citet{Zakamulin.2015}, and the out-of-sample portfolio variance. We show empirically that in most cases there exists a strong positive relation between the selection criterion within the CV and the respective out-of-sample performance measure. For instance, when the overall goal is to reduce the out-of-sample risk, then using CV with the portfolio variance as a measure of fit leads to lower risk than the original method. Similar results are documented by \citet{Liu.2014}, although he only considers the most straight-forward linear shrinkage as in \citet{Ledoit.2003, Ledoit.2004, Ledoit.2004b}. Here, we examine more recent and efficient estimation methods and identify those that can actually profit from a data-driven methodology. In detail, estimators that depend strongly on the choice of a specific tuning parameter within their derivation are more prone to be positively influenced by replacing the original solution with a data-driven one.

Our contributions to the current literature on the subject of covariance and precision matrix estimation within the portfolio optimization framework can be summarized as follows. First, we show that recent advances in methods for high-dimensional covariance estimation lead to strong improvements in the risk profile of the GMV. In this context, we emphasize the distinct and often significant outperformance of the GLASSO, which estimates sparse precision matrices by identifying pairs of zero partial auto-correlations in asset returns data. In line with the main discussion, we show that a model's outperformance in respect to out-of-sample portfolio variance does coincide with an identical objective within the CV. Although the elaborate method outlined in \citet{Ledoit.2017} is not strongly influenced by applying the CV procedure, all the other considered data-driven estimators perform better than their original counterparts. This advantage becomes even greater as the high-dimensionality of the data
increases. Considering the MSFE, the results are straightforward for all estimation methods. If an investor aims to minimize this measure, the respective data-driven estimation ought to be performed. Nonetheless, we analyze the inefficiency of the MSFE for high-dimensional asset returns' data, in particular, because of a distorted calculation of the realized covariance matrix. Finally, to test the robustness of our results, we examine the effects of the suggested methods on the GMV without short sales. As argued by \citet{Jagannathan.2003}, a constraint imposed on short sales is similar to the linear shrinkage technique since it reduces the impact of estimation errors on optimal portfolio weights. Nevertheless, we show that a CV with the portfolio variance as a selection criterion slightly improves the out-of-sample risk even in this scenario. 

The rest of  paper is organized as follows: In Section~\ref{sec:cov}, we review the considered covariance estimation methods and their properties. Section~\ref{sec:datadriven} outlines the suggested data-driven methodology in respect to its main characteristics: the cross-validation procedure, the parameter set, and the selection criteria. We describe the empirical study in Section~\ref{sec:empirical} with a strong focus on the chosen dataset, methodology and performance measures. In Section~\ref{sec:results}, we discuss the performance of classical and constrained GMV portfolios and analyze in detail the influence of data-driven estimation among all considered datasets and methods. Section~\ref{sec:conclusion} summarizes the results and concludes.

\section{Overview of the Estimation Methods}\label{sec:cov}

\subsection{Sample Covariance}\label{subsec:covsample}

The standard approach for estimating the covariance matrix of returns among researchers and practitioners is to use the sample estimator, defined as
\begin{linenomath*}
\begin{equation}\label{eq:covsample}
\widehat{{{\Sigma}}}_\mathit{S}=\frac{1}{T-1} \left({R} - \widehat{{\mu}} {1} \right)' \left({R} -  \widehat{{\mu}}{1}\right),
\end{equation}
\end{linenomath*}
where ${R} \in \mathbb{R}^{T \times n}$ is the matrix of past asset returns with $T$ observations and $n$ number of stocks, $\widehat{{\mu}} \in \mathbb{R}^{n}$ is the vector of expected returns, here estimated with the sample mean, and ${1}$ is an $n$-dimensional vector of ones. As shown by \citet{Merton.1980}, the sample covariance matrix is an asymptotically unbiased and consistent estimator of the true covariance matrix ${\Sigma}$ in terms of the Frobenius norm $||\cdot||_2$ ´when the concentration ratio $q=n/T\rightarrow 0$. For a large number of assets a concentration ratio of such magnitude is practically infeasible due to limited data availability and illiquidity issues. With a high relation of the number of assets to the sample size, also called high-dimensionality, the sample covariance and its inverse exhibit higher amount of estimation error, mainly due to the over- and underestimation of the respective eigenvalues. Moreover, for $q>1$, the sample covariance becomes singular and the inverse cannot be calculated. 

The sample estimator's instability and possible singularity in case of high-dimensionality are a problem within the optimization of global minimum-variance portfolios, where the covariance matrix and, specifically, its inverse capture the dependency between asset returns and allow for the effect of diversification as a way of reducing risk. It is then straightforward that the accuracy of optimally estimated portfolio weights is directly related to the estimator's precision. As a solution, several alternative estimators have been proposed in the literature.

\subsection{Linear Shrinkage}\label{subsec:covlin}

To produce more stable estimators of the covariance matrix, a linear shrinking procedure can be applied to the sample estimator towards a more structured target matrix $\widehat{{\Sigma}}_\mathit{T}$,
\begin{linenomath*}
\[	\widehat{{\Sigma}}_\mathit{LS}  = s\widehat{{\Sigma}}_\mathit{T} + \left(1- s\right)\widehat{{\Sigma}}_\mathit{S}\,,\]
\end{linenomath*}
where the constant $s\in[0,1]$ controls the shrinkage intensity, which is set higher the more ill-conditioned the sample estimator is and vice versa. In contrast to the unbiased, but unstable sample covariance, a structured target matrix has little estimation error but tends to be biased. As a compromise, the convex combination of both uses the bias-variance trade-off by accepting more bias in-sample in exchange for less variance out-of-sample. This idea is central to the shrinkage methodology of \citet{Stein.1956} and \citet{James.1961}. More recently, \citet{Ledoit.2004b} propose a set of asymptotically optimal shrinkage estimators and specify the identity matrix as a generic target matrix if no specific covariance structure is assumed. The respective linear shrinkage estimator is calculated as
\begin{linenomath*}
\begin{equation}\label{eq:covlwident}
\widehat{{\Sigma}}_\mathit{LW_\mathit{1}}  =  \widehat{s} \bar{\sigma} {I}_\mathit{n}  + \left(1-  \widehat{s}\right)\widehat{{\Sigma}}_\mathit{S}\,,
\end{equation}
\end{linenomath*}
where $\bar{\sigma} = \frac{1}{n}\sum^{n}_{j=1}\sigma_\mathit{jj}$ is the average of all individual sample variances and $\widehat{s}$ is an optimal shrinkage intensity parameter. However, in the context of financial time series, it is beneficial to consider target matrices with reference to the correlation structure of asset returns.

\citet{Ledoit.2004} consider identical pairwise correlations between all $n$ assets. The target matrix is therefore derived under the constant correlation matrix model of \citet{Elton.1973}, so that $\widehat{{\Sigma}}_\mathit{T}=\widehat{{\Sigma}}_\mathit{CC}$. While the variances are kept as their original sample values, the off-diagonal entries of the target matrix are estimated by assuming a constant average sample correlation $\bar{\rho}$. This results in $\widehat{{\Sigma}}_\mathit{CC,ij}=\sqrt{\widehat{\sigma}_\mathit{ii}\widehat{\sigma}_\mathit{jj}\bar{\rho}}.$ The corresponding estimator is defined as
\begin{linenomath*}
\begin{equation}\label{eq:covlwcc}
\widehat{{\Sigma}}_\mathit{LW_\mathit{CC}}  = \widehat{s} \widehat{{\Sigma}}_\mathit{CC} + \left(1-  \widehat{s}\right)\widehat{{\Sigma}}_\mathit{S}\,.
\end{equation}
\end{linenomath*}
The level of the shrinkage $\widehat{s}$ in Equations~\eqref{eq:covlwident} and \eqref{eq:covlwcc} can be obtained analytically. In particular, as shown by \citet{Ledoit.2004, Ledoit.2004b}, asymptotically consistent estimators for the optimal linear shrinkage intensity are derived under the quadratic loss function  
\begin{linenomath*}
\begin{equation}\label{eq:frobloss}
\mathcal{L}\left( \widehat{{\Sigma}}, {\Sigma}\right) = \left|\left| \widehat{{\Sigma}} - {\Sigma}\right|\right|^2_\mathit{F},
\end{equation}
\end{linenomath*}  
known as the Frobenius loss, where the covariance estimator $\widehat{{\Sigma}}$ is substituted with Equation~\eqref{eq:covlwident} or \eqref{eq:covlwcc}. The finite sample solution is found at the minimum of the expected value of the Frobenius loss, namely the mean squared error (MSE),
\begin{linenomath*}
\begin{equation}\label{eq:optlinearshrink}
\widehat{s} = \argmin_{s} E\left[\left|\left| \widehat{{\Sigma}} - {\Sigma}\right|\right|^2_\mathit{F}\right].
\end{equation}
\end{linenomath*}
The methodology behind this derivation can be applied to other shrinkage targets in a convex combination setting after an individually performed analysis and mathematical adaptation. Our data-driven implementation, however, can be implemented for any linear shrinkage without further modifications, since we do not rely on the theoretically derived shrinkage intensity; instead, we search for an optimal value using CV.

\subsection{Nonlinear Shrinkage}\label{subsec:covlwnonlin}

The nonlinear shrinkage method, first proposed by \citet{Ledoit.2012}, shrinks covariance entries by increasing small (underestimated) sample eigenvalues and decreasing large (overestimated) ones in an individual fashion. Without any assumption about the true covariance matrix, the positive-definite rotationally equivariant\footnote{This class of estimators was first introduced by \cite{Stein.1986}.} nonlinear shrinkage is based on the spectral decomposition of the sample covariance matrix and defined as
\begin{linenomath*}
\begin{equation}\label{eq:covlwnonlinear}
\widehat{{\Sigma}}_\mathit{LW_\mathit{NL}} = {V}\widehat{{\Lambda}}_\mathit{NL} {V}' \,,
\end{equation}
\end{linenomath*}
where ${V} = \left[{v}_\mathit{1}, \hdots, {v}_\mathit{n}\right]$ is the orthogonal matrix with the sample eigenvectors ${v}_\mathit{i}$ as columns and $\widehat{{\Lambda}}_\mathit{NL}$ is the diagonal matrix of the sample eigenvalues $\lambda_i$, shrunk as shown in \citet{Ledoit.2012, Ledoit.2015, Ledoit.2017, Ledoit.2018}. To find the optimal nonlinear shrinkage to the eigenvalues, \cite{Ledoit.2012} originally minimize the MSE in finite samples. Under the considered large-dimensional asymptotics from the field of Random Matrix Theory the Frobenius loss converges almost surely to a nonstochastic limit, guaranteeing the estimator's optimality.

Without going into further details, we examine the practical implementation of the nonlinear shrinkage, as demonstrated by \citet{Ledoit.2018}. The optimal solution is achieved using a nonparametric variable bandwidth kernel estimation of the limiting spectral density of the sample eigenvalues and its Hilbert transform. The speed at which the bandwidth vanishes in the number of assets $n$ can be set to $-0.2$ according to standard kernel density estimation theory \citep{Silverman.1986} or $-0.5$ following the Arrow model of \citet{Ledoit.2018b}. As a compromise between those two approaches, \citet{Ledoit.2018} suggest the value of $-0.35$. Within the suggested data-driven methodology, we aim to verify whether this exact choice of the kernel bandwidth's speed is crucial for the estimator's efficiency and whether applying the suggested CV technique can improve the out-of-sample performance.

\subsection{Approximate Factor Model}\label{subsec:covafm}
The previously outlined methods for improved high-dimensional covariance estimation do not assume any structural knowledge about the covariance matrix and regularize only the sample eigenvalues $\lambda_i$. An underlying structure could be established by regularizing the sample eigenvectors ${v}_i$, for example if the covariance matrix itself is assumed to be sparse \citep[see, e.g.,][]{Bickel.2008, Cai.2011}. Unfortunately, this is not appropriate for financial time series because of the presence of common factors \citep{Fan.2013}. However, if there is only conditional sparsity, the covariance matrix of investment returns can be estimated using factor models given by
\begin{linenomath*}
\[\widehat{{\Sigma}}_\mathit{FM} = {B}\widehat{{\Sigma}}_\mathit{F}{B}' + \widehat{{\Sigma}}_\mathit{u}\,,\]
\end{linenomath*}
where ${\Sigma}_\mathit{F}$ is the sample covariance matrix of the common factors and $\widehat{{\Sigma}}_\mathit{u}$ is the residuals covariance matrix.\footnote{Following this definition and assuming $K$ common factors with $K<n$, a covariance matrix estimator based on factor models only needs to estimate $K(K+1)/2$ covariance entries and is thus more stable.} One disadvantage of such exact factor models is the strong assumption of no correlation in the error terms across assets; that is, the error covariance matrix $\widehat{{\Sigma}}_\mathit{u}$ is assumed to contain only the sample variances of the residuals. Therefore, possible cross-sectional correlations are neglected after separating the common present factors \citep{Fan.2013}. Instead, approximate factor models allow for off-diagonal values within the error covariance matrix. The POET estimator is one of the most recent and efficient estimators from this branch of research. Using the close connection between factor models and the principal component analysis, \citet{Fan.2013} infer the necessary factor loadings by running a singular value decomposition on the sample covariance matrix as
\begin{linenomath*}
\[
{\Sigma}_\mathit{S}=\sum^{K}_{i=1}\lambda_\mathit{i} {v}_\mathit{i} {v}_\mathit{i}' + \sum^{n}_{i=K+1}\lambda_\mathit{i} {v}_\mathit{i} {v}_\mathit{i}'.
\]
\end{linenomath*}
The covariance, formed by the first $K$ principal components, contains most of the information about the implied structure. The rest is assumed to be an approximately sparse matrix, estimated by applying an adaptive thresholding procedure \citep{Cai.2011} with a threshold parameter $c$.\footnote{For the operational use of POET, the threshold value $c$ needs to be determined, so that the positive-definiteness of $\widehat{{\Sigma}}^{c}_\mathit{u,K}$ is assured in finite samples. The choice of $c$ can therefore occur from a set, for which the respective minimal eigenvalue of the errors' covariance matrix after thresholding is positive. The minimal constant $c$ that guarantees positive-definiteness is then chosen. For more details, see, \citet{Fan.2013}.} As a result, the POET estimator becomes
\begin{linenomath*}
\begin{equation}\label{eq:covPOET}
{\Sigma}_\mathit{POET}=\sum^{K}_{i=1}\lambda_\mathit{i} {v}_\mathit{i} {v}_\mathit{i}' + \widehat{{\Sigma}}^{c}_\mathit{u,K} \,.
\end{equation}
\end{linenomath*}
As argued by \citet{Fan.2013}, for high-dimensional asset returns with a sufficiently large $n\rightarrow \infty$, the number of factors $K$ can be inferred from the data. A consistent data-driven estimator for $K$ is 
\begin{linenomath*}
\begin{equation}\label{eq:covPOETK}
\widehat{K}= \argmin_{0\leq k \leq k_\textrm{max}} \log\left(\frac{1}{nT}\left|\left| {R} - \frac{1}{T}{R} {F}_\mathit{k} {F}_\mathit{k}' \right|\right|^{2}_\mathit{F}\right) + k g\left(T,n\right),
\end{equation}
\end{linenomath*}
where $k_\textrm{max}$ is the predefined maximum number of factors, ${R}$ is the matrix of asset returns with a sample covariance matrix ${\Sigma}_\mathit{S}$, ${F}_\mathit{k}$ is a $T\times k$ matrix with columns the eigenvectors, corresponding to the $k$ largest eigenvalues of ${\Sigma}_\mathit{S}$, and $ g\left(T,n\right)$ is a penalty function of the type, introduced by \citet{Bai.2002}. In this study we further examine whether the proposed CV approach can select optimal values for $K$ by considering the out-of-sample performance measure of interest as a selection criterion.

\subsection{Graphical Model}\label{subsec:covglasso}

A proper estimation of the covariance matrix of returns is crucial in a portfolio optimization context, since its inverse ${\Theta}={\Sigma}^{-1}$ is the direct input parameter necessary for exploiting diversification effects upon optimization. Instead of imposing a factor structure on the covariance matrix with a sparse error covariance as in POET, sparsity in the precision matrix can be a valid approach for reducing estimation errors, especially in the case of conditional independence among asset pairs \citep{Fan.2016}. In detail, the entry ${\Theta}_\mathit{i,j}=0$ if and only if asset returns ${r}_i$ and ${r}_j$ are independent, conditional on the other assets in the investment universe. Since graphical models are used to describe both the conditional and unconditional dependence structures of a set of variables, the estimation of ${\Theta}$ is closely related to graphs under a Gaussian model. The identification of zeros in the inverse can be performed with the Gaussian graphical model, since within the Markowitz portfolio optimization framework asset returns are assumed to follow a multivariate normal distribution.\footnote{This idea was first proposed by \cite{Dempster.1972} with the so-called covariance selection model.}

One of the most commonly used methods for inducing sparsity on the precision matrix is by penalizing the maximum-likelihood. For i.i.d. ${R}$ with ${R}\sim\mathcal{N}\left({0}, {\Sigma}\right)$, the Gaussian log-likelihood function is given by
\begin{linenomath*}
\begin{equation}\label{eq:loglike}
\mathcal{L}\left({\Theta}\right)= \log\left|{\Theta}\right| - \text{tr}\left( \widehat{{\Sigma}}_\mathit{S}{\Theta}\right),
\end{equation}
\end{linenomath*}
where $|\cdot|$ denotes the determinant and $\textrm{tr}(.)$ the trace of a matrix. Maximizing Equation~\eqref{eq:loglike} alone yields the known maximum-likelihood estimator for the precision matrix $\widehat{{\Theta}}_\mathit{S}$, which suffers from high estimation error in case of high-dimensionality. To reduce such errors, the maximum log-likelihood function can be penalized by adding a lasso penalty \citep{Tibshirani.1996} on the precision matrix entries as
\begin{linenomath*}
\begin{equation}\label{eq:covglassologlike}
\mathcal{L}\left({\Theta}\right)= \log\left|{\Theta}\right| - \text{tr}\left( \widehat{{\Sigma}}_\mathit{S}{\Theta}\right) - \rho \left|\left|{\Theta}^{-}\right|\right|_1,
\end{equation}
\end{linenomath*}
where $\left|\left|{\Theta}^{-}\right|\right|_1$ is the $L_1$-norm (the sum of the absolute values) of the matrix ${\Theta}^{-}$, an $n\times n$ matrix with the off-diagonal elements, equal to the corresponding elements of the precision matrix ${\Theta}$ and the diagonal elements equal to zero.\footnote{This insures that no penalty is applied to the asset returns' sample variances.} Furthermore, $\rho$ is a penalty parameter that controls the sparsity level, with higher $\rho$ values leading to a larger number of off-diagonal zero elements within the resulting estimator. 

The penalized likelihood framework for a sparse graphical model estimation was first proposed by \citet{Yuan.2007}, who solve Equation~\eqref{eq:covglassologlike} with an interior-point method. \citet{Banerjee.2008} show that the problem is convex and solve it for ${\Sigma}$ with a box-constrained quadratic program. To date, the fastest available solution for the sparse graphical model in Equation~\eqref{eq:covglassologlike} is reached with the GLASSO algorithm, developed by \citet{Friedman.2008} and later improved by \citet{Witten.2011}. They demonstrate that the above formulation is equivalent to an N-coupled lasso problem and solve it using a coordinate descent procedure. 

In addition to a well-performing algorithm, the value of $\rho$ is necessary for calculating the optimal GLASSO estimator. For this purpose, \citet{Yuan.2007} suggest using the Bayesian Information Criterion (BIC), defined for each $\rho$ as 
\begin{linenomath*}
\begin{equation}\label{eq:covglassoBIC}
BIC(\rho) = - \log\left|\widehat{{\Theta}}_\mathit{\rho}\right| + \text{tr}\left( \widehat{{\Sigma}}_\mathit{S}\widehat{{\Theta}}_\mathit{\rho}\right) + \frac{\log(T)}{2} \sum^{n}_{i=1, i\neq j}\sum^{n}_{j = 1, j \neq i}\mathbbm{1}_{\{\widehat{{\Theta}}_\mathit{\rho,ij}\neq0\}},
\end{equation}
\end{linenomath*}
where the indicator function $\mathbbm{1}_{\{\widehat{{\Theta}}_{\rho,ij}\neq0\}}$ counts the number of nonzero off-diagonal elements in the estimated precision matrix. The value of $\rho$, corresponding to the lowest BIC, is chosen as the optimal lasso penalty parameter. The choice of the BIC as a selection criterion for $\rho$ is further justified by the relation between the penalized problem in Equation~\eqref{eq:covglassologlike} and the model selection criteria \citep{Goto.2015}. Although \citet{Yuan.2007} argue that a CV procedure for an optimal lasso penalty can yield better out-of-sample results, the existing financial applications estimate $\rho$ only once in-sample.\footnote{\citet{Goto.2015} induce sparsity to enhance robustness and lower the estimation error within portfolio hedging strategies, \citet{Brownlees.2018} develop a procedure called ``realized network'' by applying GLASSO as a regularization procedure for realized covariance estimators, and \citet{Torri.2019} analyze the out-of-sample performance of a minimum-variance portfolio, estimated with GLASSO.} By contrast, next to such conservative approach, we consider the superiority of data-driven methods in the context of lasso regularization and perform additionally a multi-fold CV with risk-related selection criteria. The exact methodology is described in the next section.

\section{Data-Driven Methodology}\label{sec:datadriven}

Each of the outlined covariance estimators includes an exogenous or data-dependent parameter. The linear shrinkage estimators in Equations~\eqref{eq:covlwident} and \eqref{eq:covlwcc} are calculated with an optimal shrinkage intensity $\widehat{s}$. For the more general nonlinear shrinkage \citet{Ledoit.2017} set the kernel bandwidth's speed at $-0.35$ as the average of two recognized approaches. The approximate factor model, the POET estimator by \citet{Fan.2013}, deals with an unknown number of factors $K$, which are identified by minimizing popular information criteria. Finally, the GLASSO estimator proposed by \citet{Friedman.2008} needs an optimal choice for the penalty parameter $\rho$, often estimated by minimizing the BIC in-sample. To clarify our analysis, we refer to these estimation methods as `original'. In addition, we adopt a nonparametric technique, a multi-fold CV, to identify the necessary parameter for each estimation method in a data-driven way. Instead of relying on pre-specified assumptions and deriving corresponding solutions individually, we perform a grid search over a domain of values and find the best possible parameter for two exemplary out-of-sample statistics.

\subsection{Parameter Set}\label{subsec:param}

To employ a data-driven choice, we first need to specify a domain of possible values for the necessary parameters that should be selected within the CV procedure. For this purpose we create a sequence (or grid) of arbitrary parameters $\delta \in \Delta$ for each covariance model. Depending on the chosen length of the sequence, the CV can be computationally time-consuming. Since the choice of this sequence is crucial for the out-of-sample efficiency of the data-driven methodology, the domain of possible parameters has to be individually evaluated for each estimation method by considering the trade-off between desired precision and computing time. Subsection~\ref{subsec:data} outlines the examined sequences for the considered covariance estimation methods.

\subsection{Cross-Validation Procedure}\label{subsec:cv}

The CV is a model validation technique designed to assess how an estimated model would perform on an unknown dataset. To evaluate the model accuracy, the available dataset is repeatedly split into a training and a testing subset in a rolling-window fashion \citep[see, e.g.,][]{Refaeilzadeh.2009, Arlot.2010}. For instance, in the case of an $m$-fold CV, a dataset with $\tau$ observations is split into $m$ equal parts. The first rolling-window then uses as a training dataset the first fold consisting of the first $\nu < \tau$ observations ordered by time. Upon this, the consecutive $\upsilon$ observations are used to validate the performed estimation as a test dataset. This is iteratively done $m$ times by shifting the training window by $\upsilon$ observations and, therefore, maintaining the chronological order within the data.

In our setting, for each of the pre-defined parameters we successively use the training data to calculate a covariance matrix estimator $\widehat{{\Sigma}}_\mathit{t,\delta}$ for a test dataset $t$ and a specific parameter $\delta$.\footnote{For clarity in the notation, we do not differentiate between covariance estimators. The procedure is applied to all methods equally.} During the following validation stage, we must set selection criteria, also referred to as measures of fit, to identify which parameter performs best. In this study, we investigate two common objectives within the field of portfolio risk minimization. 

As often argued, the squared forecasting error (SFE) or, as defined in Section~\ref{sec:cov}, the Frobenius loss, is minimized to find a covariance estimator with the least forecasting error \citep[see, e.g.,][]{Zakamulin.2015}. Specifically, we first calculate a realized covariance matrix for the test dataset with
\begin{linenomath*}\[{\Sigma}_\mathit{t}= \left({R}_\mathit{t}- \widehat{{\mu}}_\mathit{t} {1} \right)' \left({R}_\mathit{t} - \widehat{{\mu}}_\mathit{t} {1} \right),\]\end{linenomath*}      
where ${R}_\mathit{t}\in \mathbbm{R}^{\upsilon \times n}$ are the asset returns from the test dataset and $\widehat{\mu}_\mathit{t}$ is the vector of average returns for the testing period consisting of $\upsilon$ observations. Then, we find the corresponding SFE as
\begin{linenomath*}\[\left|\left|\widehat{{\Sigma}}_\mathit{t,\delta} - {\Sigma}_\mathit{t}\right|\right|^2_\mathit{F}.\]\end{linenomath*}
This procedure is repeated $m$ times, so that we end up with $m$ SFE values for each $\delta$. From the parameter set we then choose this $\delta$ for which the average (over all $m$ iterations) SFE is minimized. In our empirical study, the data-driven estimation method with the SFE as a measure of fit is referred to as CV1.  

Instead of the SFE, within a portfolio optimization framework, one is generally more interested in whether a covariance estimator leads to lower out-of-sample risk of the optimal portfolio \citep[see, e.g.,][]{Liu.2014, Ledoit.2017, Engle.2019}. To incorporate and later investigate this concept, as our second scenario (CV2), we minimize the out-of-sample portfolio variance. In detail, with the covariance matrix $\widehat{{\Sigma}}_\mathit{t, \delta}$, previously estimated with the training data, we calculate the optimal weights $\widehat{{w}}_\mathit{t,\delta}$ for a portfolio of our choice (e.g., the GMV). This then allows us to calculate the respective portfolio returns throughout the testing period with $\upsilon$ observations as
\begin{linenomath*}\[{r}^p_\mathit{t, \delta}=\widehat{{w}}_\mathit{t,\delta}'{R}_\mathit{t}.\]\end{linenomath*}
This procedure is repeated $m$ times, so that we end up with $m$ portfolio return vectors for each $\delta$. From the parameter set, we then choose this $\delta$ for which the empirical variance (over all $m$ iterations) of those portfolio out-of-sample returns is minimized.

By applying different measures of fit within the data-driven methodology we explicitly address the importance of aligned selection criteria for the out-of-sample performance of each covariance estimation method. Moreover, we aim to verify whether the estimation of covariance parameters with a multi-fold CV yields better results out-of-sample than the original models.

\section{Empirical Study}\label{sec:empirical}

To exploit the above considerations, we perform an extensive empirical study of the suggested covariance estimation methods within a high-dimensional portfolio optimization context. For this purpose, we create GMV portfolios with and without short sales and evaluate their out-of-sample performance for a range of commonly used measures. We additionally compare the theoretical covariance parameters with their calibrated equivalents. The exact empirical construct is elaborated on in the following subsections.

\subsection{Model Setup}\label{subsec:model}

For the empirical study, we focus on the GMV portfolio. The optimal weights for an investment period $t$ are determined by minimizing the portfolio variance as
\begin{linenomath*}
\begin{equation}\label{eq:gmv}
\begin{aligned}
\widehat{{w}}_\mathit{t} = \argmin_{{w}} \quad & {w}' \widehat{{\Sigma}}_\mathit{t} {w}\\
\textrm{s.t.} \quad &  {1}_n' {w} = 1\,,
\end{aligned}
\end{equation}
\end{linenomath*}
where ${1}_n$ is an n-dimensional vector of ones and $\widehat{{\Sigma}}_\mathit{t}$ is an arbitrary covariance matrix estimator for the investment period $t$. This formulation has the analytical solution $\widehat{{w}}_\mathit{t}=\frac{\widehat{{\Sigma}}^{-1}_\mathit{t}{1}_n}{{1}_n'\widehat{{\Sigma}}^{-1}_\mathit{t}{1}_n}.$
Furthermore, we consider a GMV portfolio with an imposed constraint on the weights (GMV-NOSHORT),
\begin{linenomath*}
\begin{equation}\label{eq:gmvnoshort}
\begin{aligned}
\widehat{{w}}_\mathit{t} = \argmin_{{w}} \quad & {w}' \widehat{{\Sigma}}_\mathit{t} {w}\\
\textrm{s.t.} \quad &  {1}_n' {w} = 1 \textrm{ and } {w}\geq {0}\,,
\end{aligned}
\end{equation}
\end{linenomath*}
which is a quadratic optimization problem with linear constraints that can be solved with every popular quadratic optimization software.\footnote{The measure of fit within CV2, the portfolio variance, depends on the estimated optimal weights. When short sales are allowed, we use the solution of Equation~\eqref{eq:gmv} to calculate the portfolio variance as in Subsection~\ref{subsec:cv}. For the case of GMV-NOSHORT, we solve Equation~\eqref{eq:gmvnoshort} within the CV.} As discussed by recent literature \citep[see, e.g.,][]{Jagannathan.2003, DeMiguel.2009}, the introduction of a short-sale constraint is not only practically relevant because of common fund rules or budget constraints for individual investors. It moreover limits the estimation error in the portfolio weights. As a consequence, we would expect a slightly reduced effect of the estimation methods' efficiency in respect to the out-of-sample performance. The empirical analysis of GMV-NOSHORT portfolios thus aims to complement our study and to ensure the practical reproducibility and relevance of our results.

\subsection{Data and Methodology}\label{subsec:data}

To test the performance of the proposed covariance matrix estimation methods, we utilize four S\&P\,500 related datasets, which differ only in the number of assets used: 50, 100, 200 and 250 stocks. Throughout our study the datasets are referred to as 50SP, 100SP, 200SP and 250SP, respectively. Choosing datasets with different quantities of assets, we aim to study the behavior of covariance estimators when the concentration ratio becomes increasingly large in-sample. As common within the research on such high-dimensional covariance estimation \citep[see, e.g.,][]{Fan.2013, Ledoit.2017, Engle.2019}, we consider daily prices.\footnote{The price history originates from the Thomson Reuters EIKON database.} For our out-of-sample analysis we use daily returns, starting on 01/01/1990 and ending on 12/31/2018. Overall, our data include observations for $T=348$ months (or 7306 days) per asset.

To ensure the stability in our results, we randomly select the necessary number of stocks among all the companies that have survived throughout the investigated period and keep them as the investment universe for our empirical study. Since the chosen datasets consist of individual stocks, the adopted investment strategies can be recreated easily and cost efficiently in practice by simply buying or selling the respective amount of stocks.

To evaluate the out-of-sample performance of the constructed portfolios and, implicitly, the covariance estimation methods, we adopt a rolling-window study with an in-sample period of two years, $\tau=24$ months (or roughly 504 days), and an out-of-sample period from 01/01/1992 to 12/31/2018, resulting in $T-\tau=324$ months (or 6801 days) out-of-sample portfolio returns. Similarly to the original studies on the reviewed covariance estimation methods \citep{Fan.2013, Ledoit.2017, Engle.2019}, we employ a monthly rebalancing strategy, since this is more cost efficient and common in practice. Within each rolling-window step, the covariance matrix of asset returns for the investment month $t$ is estimated at the end of month $t-1$ using approximately the most recent 504 daily in-sample observations.  

In our empirical study, the sample covariance estimator serves as a benchmark to the high-dimensional estimation methods and considered data-driven adjustments in terms of out-of-sample risk. For the application of the nonlinear shrinkage, we use the MATLAB-code provided by Ledoit and Wolf.\footnote{\url{https://www.econ.uzh.ch/en/people/faculty/wolf/publications.html}.}. The POET estimator is calculated using the R-package \verb|POET| provided by \citet{Fan.2013}. As suggested by the authors, we adopt a soft-thresholding rule as well as a data-driven derivation of the number of factors and the thresholding constant for the original version of the method. Finally, the GLASSO estimator is calculated with the algorithm provided by \cite{Friedman.2008} within the R-package \verb|glasso| with no penalty on the diagonal elements. 

In addition to the models in Section~\ref{sec:cov}, we calculate the data-driven estimators as in Section~\ref{sec:datadriven} by applying an $m$-fold CV. To calculate the selection criteria for the respective CV methods, we choose $m=12$ and therefore divide the in-sample observations into a training sample of 12 months (or 252 days) and a testing sample of one month (or 21 days). With this construction, we replicate the proposed monthly rebalancing strategy inside the performed CV. As introduced in Subsection~\ref{subsec:param}, we additionally need to define a set of parameters for each covariance estimation method. 

Since both linear shrinkage methods in Equation~\eqref{eq:covlwident} (LW\textsubscript{1}) and Equation~\eqref{eq:covlwcc} (LW\textsubscript{CC}) represent the weighted average between the sample and a target covariance matrix, we define a parameter set $\Delta$ of G shrinkage intensities, such that $\Delta_\mathit{1}=\Delta_\mathit{CC}=\left(\delta_1, \delta_2, \hdots, \delta_G\right)\in\left[0,1\right]$. Considering the reasoning in Subsection~\ref{subsec:covlwnonlin}, for the nonlinear shrinkage estimator in Equation~\eqref{eq:covlwnonlinear} (LW\textsubscript{NL}), we set the kernel bandwidth's speed to lie between $-0.2$ and $-0.5$ with $\Delta_\mathit{NL}=\left(\delta_1, \delta_2, \hdots,  \delta_G\right)\in\left[-0.2,-0.5 \right]$. Since the accuracy of a data-driven estimation depends on the number of examined parameters, with more parameters allowing for finer results, we consider a linear grid of $G=1000$ equidistant values in the above cases. Furthermore, for the POET estimator, we consider $\Delta_\mathit{POET}=\left(\delta_1, \delta_2, \hdots, \delta_G\right)\in \{1,2,\ldots,6\}$. For the GLASSO estimator, we follow \citet{Friedman.2008} and choose a sequence of penalty parameters $\rho$, derived from the training data. Specifically, we define a logarithmic sequence $10^{\log_{10}(k(x, e, u, G))}$ as our $\rho$-generating function, where $k(x, e, u, G)=(x-1)\cdot\frac{e-u}{G-1}+u$ with $G=50$ number of parameters in the sequence, $u$ being the maximal absolute value of the sample covariance matrix, estimated with the training dataset, and $e=0.01u$. 

After calculating all the possible combinations of original and data-driven estimators within the validation subset, we choose an optimal parameter for each covariance estimation method, as outlined in Subsection~\ref{subsec:cv}, and use all the in-sample data to estimate the covariance matrix for the next investment month. Since the reviewed estimation methods and our data-driven methodology do not model time-dependency in the covariance matrix, we set $\widehat{{\Sigma}}_\mathit{t}=\widehat{{\Sigma}}_\mathit{t-1}$. We use $\widehat{{\Sigma}}_\mathit{t}$ to find the optimal weights $\widehat{{w}}_\mathit{t}$, as in Equations~\eqref{eq:gmv} and \eqref{eq:gmvnoshort}. With these weights, we calculate the out-of-sample portfolio returns for each model in $t$. This procedure is repeated multiple times until the end of our investment horizon.

Overall, our study covers 17 different portfolios for each scenario with and without short sales. First, we include the equally-weighted portfolio, hereafter also referred to as the Naive portfolio. This strategy implies an identity covariance matrix and hence, does not include any estimation risk \citep{DeMiguel.2009}. In addition to the Naive strategy, which is a standard benchmark when comparing induced transaction costs, we build a GMV portfolio with the sample covariance matrix estimator, which serves as a benchmark for the out-of-sample risk. For each of the five high-dimensional covariance estimation methods discussed in Section~\ref{sec:cov}, we construct portfolios with the original and calibrated parameters, resulting in three versions for each estimation methodology. All these portfolios are evaluated with the performance measures, presented in the following subsection. 

\subsection{Performance Measures}\label{subsec:perf}

To evaluate the out-of-sample performance of each covariance matrix estimation method, we report different performance measures for the estimator's efficiency and the risk profile as well as the allocation properties of the corresponding GMV and GMV-NOSHORT portfolios. First, we calculate the MSFE as
\begin{linenomath*}
\begin{equation}\label{eq:MSFE}
\text{MSFE}=\frac{1}{T-\tau}\sum^{T-\tau}_{t=\tau}\sum^{n}_{i=1}\sum^{i}_{j=1}\left({\Sigma}_{t, ij} - \widehat{{\Sigma}}_{t, ij}\right)^2,
\end{equation}
\end{linenomath*}
where $\widehat{{\Sigma}}_{t, ij}$ is the covariance matrix estimator and ${\Sigma}_{t, ij}$ is the realized covariance for month $t$. The MSFE is frequently used to measure the forecasting power of an estimation method. To avoid double accounting for forecasting errors, we exclude the lower triangular part of both matrices from the calculation.

Considering the nature of minimum-variance portfolios as risk-reduction strategies, we are especially interested in the out-of-sample SD as a performance indicator. We calculate the standard deviation (SD) of the 6801 out-of-sample portfolio returns and multiply by $\sqrt{252}$ to annualize it. For a more detailed analysis of the out-of-sample risk of the constructed portfolios and therefore, implicitly, covariance estimation methods, we perform the two-sided Parzen Kernel HAC-test for differences in variances, as described by \citet{Ledoit.2008} and \citet{Ledoit.2011}, and report the corresponding significance levels. Since we utilize daily returns, a sufficient number of observations is available and a bootstrap technique is not essential.\footnote{For the sake of completeness, we have also performed a block bootstrap as in \citet{Ledoit.2011}. The corresponding significant values are comparable to those from the HAC test and are therefore not reported.} Since the MSFE is closely related to the SFE optimality criterion, as within the CV1 method, we expect the respectively optimized covariance estimators to exhibit a lower MSFE than their original versions. Moreover, a data-driven estimation with the CV2 approach, based on minimizing the portfolio variance, is expected to result in a lower out-of-sample SD.

In practice investors need to additionally address the problem of high transaction costs; hence, they prefer a more stable allocation for an optimal portfolio strategy. Therefore, as a proxy for occurring transaction costs, we analyze the average monthly turnover, defined as
\begin{linenomath*}
\begin{equation}\label{eq:turnover}
\textrm{Turnover}=\frac{1}{T-\tau-1}\sum^{T-\tau-1}_{t=\tau}\left|\left|\widehat{{w}}_{t+1}-\widehat{{w}}^{+}_{t}\right|\right|_1,
\end{equation}
\end{linenomath*}
where $||\cdot||_1$ denotes the $\ell_1$-norm of a vector as the sum of its absolute values and $\widehat{{w}}^{+}_{t}$ denotes the portfolio weights at the end of the investment month $t$, scaled back to one. The turnover rate is calculated as the averaged sum of absolute values of the monthly rebalancing trades across all $n$ assets and over all investment dates $T-\tau-1$. The next section reports the detailed out-of-sample performance analysis and empirical results.

\section{Empirical Results}\label{sec:results}

In the empirical part to this paper we compare how the original and data-driven methods for estimating the covariance matrix of returns affect the out-of-sample performance of GMV portfolios with and without short positions. This section examines the out-of-sample properties of the three estimation methodologies (original, CV1 and CV2).

\subsection{Optimal Parameters}

\begin{figure}[h!]
	\begin{center}
		\resizebox*{\textwidth}{!}{\includegraphics{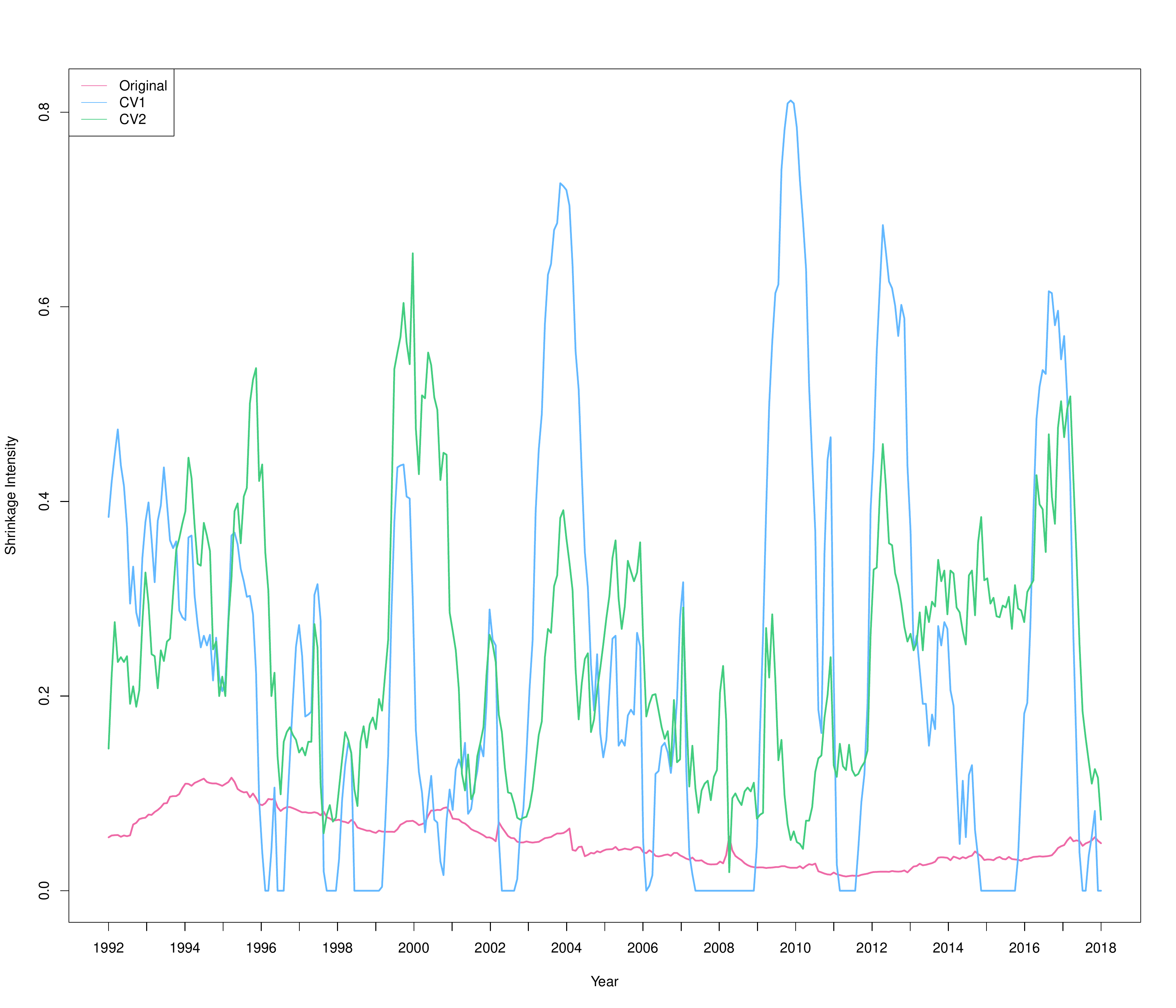}}
		\caption{Optimally selected shrinkage intensity for the original, CV1, and CV2 estimations for LW\textsubscript{1} and the 50SP dataset.}\label{fig:covparameterLW1}
	\end{center}
\end{figure}

Figure~\ref{fig:covparameterLW1} exemplary displays the selected linear shrinkage intensities for the original as well as CV1- and CV2-based LW\textsubscript{1} estimation methods in the case of 50 stocks.\footnote{Figure~\ref{fig:covparameters} shows the evolution of the selected parameters for the remaining covariance estimation methods in the case of the 50 considered stocks. The other three datasets produce similar results. To our surprise, the code for the POET estimator provided by \citet{Fan.2013} produces a consistent $K=2$ number of factors throughout the observation period and for all four datasets.} The trend of the optimal linear shrinkage intensities shows that the original approach of \citet{Ledoit.2004b} is less reactive to changes in asset returns than our CV methodologies. The strong fluctuation in the selected shrinkage intensity for CV1 and CV2 results from their data-driven nature which implies fast adaptation to potentially changing market conditions. Nevertheless, such volatility in the parameter estimation could have negative effects on the out-of-sample properties of the corresponding estimators and thus, the estimated portfolios (e.g., in terms of turnover or an overall risk level). Therefore, the sole observation of the chosen parameters cannot lead to a clear conclusion on whether the CV technique enhances the covariance matrix estimator and the respective portfolio performance. In the following subsections we examine this behavior for the GMV portfolios with short sales.

\subsection{GMV with short sales}\label{subsec:resultsgmv}

Table~\ref{table:perf} presents the central results of our empirical analysis on the GMV portfolio. The columns show the investment universes with 50, 100, 200 and 250 randomly chosen S\&P 500 stocks as well as the three performance measures MSFE, SD and average monthly turnover rate (TO). The rows indicate the portfolio strategies based on the covariance estimation. While the original estimators are noted only by the respective name of the estimation method, the endings CV1 and CV2 represent the data-driven approaches, as explained in the previous sections.
	
\begin{sidewaystable}[ph!]
	\caption{Performance of GMV portfolios across different estimators and datasets.}\label{table:perf}
	\resizebox{\textwidth}{!}{%
		\begin{threeparttable}
			\begin{tabular}{l*{3}c@{\hskip 0.3in}*3{c}@{\hskip 0.3in}*3{c}@{\hskip 0.3in}*3{c}}\toprule
				\multicolumn{3}{r}{50SP}  & \multicolumn{3}{r}{100SP} & \multicolumn{3}{r}{200SP} & \multicolumn{3}{r}{250SP}  \\ \midrule
				& MSFE & SD & TO & MSFE & SD & TO & MSFE & SD & TO & MSFE & SD & TO \\ 
				& & & & & & & & & & & & \\
				Naive &  & 0.1887 & 0.0596 &  & 0.1811 &  0.0587  &  & 0.1852 & 0.0601 &  & 0.1858 & 0.0592 \\ 
				Sample & 0.0974 & 0.1265 & 0.3426 & 0.3224 & 0.1163 & 0.6578 & 1.3344 & 0.1180 &  1.5118  & 1.9834 & 0.1216 & 2.0891 \\ 
				& & & & & & & & & & & & \\
				LW\textsubscript{1} & 0.0970 & 0.1250 & 0.2945 & 0.3211 & 0.1132 & 0.5229 & 1.3297 & 0.1092 & 0.9742 & 1.9768 & 0.1074 & 1.2052 \\ 
				LW\textsubscript{1}-CV1 & \underline{\textbf{0.0887}} & 0.1263 &  0.2623  & \underline{\textbf{0.2911}} & 0.1145 & 0.4527  & \underline{\textbf{1.1980}} & 0.1112 & 0.8662 & \underline{\textbf{1.7720}} & 0.1120 & 1.3532 \\ 
				LW\textsubscript{1}-CV2 & 0.0953 & \textbf{0.1248} & \textbf{0.2236} & 0.3074 & \textbf{0.1112} & \underline{\textbf{0.3017}} & 1.2621 & \textbf{0.1053} & \textbf{0.4229} & 1.8644 & \textbf{0.1019} & \textbf{0.4998}  \\ 
				& & & & & & & & & & & & \\
				LW\textsubscript{CC} & 0.0972 & 0.1244 &  0.2773 & 0.3218 & 0.1130 & 0.5097 & 1.3336 & 0.1092 & 1.0257 & \textbf{1.9819} & 0.1060 & 1.2435 \\ 
				LW\textsubscript{CC}-CV1 & \textbf{0.0967} & 0.1261 & \textbf{0.2469} & \textbf{0.3214} & 0.1146 & 0.4365  & \textbf{1.3333} & 0.1098 & 0.8360 & 1.9825 & 0.1083 & 1.4716 \\ 
				LW\textsubscript{CC}-CV2 & 0.0971 & \textbf{0.1244} & 0.2474 & 0.3218 & \textbf{0.1120} & \textbf{0.3699} & 1.3508 & \textbf{0.1076} & \textbf{0.5502} & 2.0050 & \textbf{0.1025} & \textbf{0.6201} \\ 
				& & & & & & & & & & & & \\
				LW\textsubscript{NL} & 0.0974 & 0.1246 & \textbf{0.2753} & 0.3222 & 0.1118 & \textbf{0.4336} & 1.3329 & 0.1053 & \textbf{0.6712} & 1.9811 & 0.1020 & \textbf{0.7307} \\ 
				LW\textsubscript{NL}-CV1 & \textbf{0.0973} & 0.1247 &  0.2817 & \textbf{0.3221} & 0.1118 & 0.4406 & \textbf{1.3322} & \textbf{0.1053} & 0.6798 & \textbf{1.9810} & \textbf{0.1020} & 0.7396  \\ 
				LW\textsubscript{NL}-CV2 & 0.0974 & \textbf{0.1246} & 0.2817 & 0.3221 & \textbf{0.1118} & 0.4393  & 1.3329 & 0.1053 & 0.6774 & 1.9811 & 0.1021 & 0.7397 \\ 
				& & & & & & & & & & & & \\
				POET & 0.0975 & \textbf{0.1238} & \textbf{0.2855} & 0.3225 & 0.1119 & \textbf{0.5233} & 1.3336 & \textbf{0.1070} &  \textbf{1.1009} & 1.9843 & \textbf{0.1041} & \textbf{1.4242} \\ 
				POET-CV1 & \textbf{0.0974} & 0.1240 &  0.2974 & \textbf{0.3223} & 0.1115 &  0.5457 & \textbf{1.3333} & 0.1072 & 1.1384 & \textbf{1.9831} & 0.1054 & 1.5588 \\ 
				POET-CV2 & 0.0974 & 0.1239 &  0.3062 & 0.3224 & \textbf{0.1114} &  0.5456 & 1.3343 & 0.1072 & 1.1879 & 1.9836 & 0.1044 & 1.5188 \\ 	
				& & & & & & & & & & & & \\
				GLASSO & 0.0974 & 0.1242 &  0.4552 & 0.3222 & 0.1105 & 0.5522 & 1.3303 & 0.1054 &  0.5385 & 1.9668 & 0.1017 &  0.6117 \\ 
				GLASSO-CV1 & \textbf{0.0941} & 0.1263 & 0.2299 & \textbf{0.3072} & 0.1147 & \textbf{0.3148} & \textbf{1.2591} & 0.1124 &  \underline{\textbf{0.3410}} & \textbf{1.8577} & 0.1099 & \underline{\textbf{0.3603}} \\ 
				GLASSO-CV2 & 0.0967 & \underline{\textbf{0.1221}} &  \underline{\textbf{0.2178}}  & 0.3180 & \underline{\textbf{0.1088}} & 0.3164 & 1.3159 & \underline{\textbf{0.1046}} & 0.4042 & 1.9579 & \underline{\textbf{0.1008}} & 0.4150 \\ 
				\bottomrule
			\end{tabular}
			\begin{tablenotes}
				\item This table reports the annualized out-of-sample SD and average monthly turnover (TO) of the GMV portfolios as well as the monthly MSFE of the respective covariance estimators across all the considered datasets with 50, 100, 200 and 250 stocks, respectively. Since the Naive portfolio strategy does not require a covariance estimator per definition, no values are reported for the MSFE. We report the lowest MSFE, SD, and TO for each estimation method in bold. The best results in terms of the MSFE and SD for each dataset are underlined. We additionally underline the lowest TO, excluding the Naive portfolio.
			\end{tablenotes}
		\end{threeparttable}
	}
\end{sidewaystable}

The compact representation of the results allows us to observe that in the case of enhanced covariance estimators the annualized SD declines as more assets are included in the GMV portfolio. This is easily explained by the known power of diversification -- the desirable effect of including more stocks in a portfolio. Not surprisingly however, the estimation error with the sample covariance estimator diminishes the positive diversification effect, as shown by the increase in out-of-sample risk for the scenarios with 200 and 250 stocks. All the efficient covariance estimation methods perform better than the sample estimator in terms of out-of-sample risk for all the datasets, with larger deviations for a higher concentration ratios.\footnote{Figure~\ref{fig:sdcomparison} provides a more visually attractive summary of the results for the GMV portfolios in terms of out-of-sample risk. For those estimation methods more susceptible to the tuning parameter, the differences in SD between the original, CV1, and CV2 methods are more pronounced.}

More importantly, we can clearly detect the positive effect of the appropriate choice of selection criterion for determining the necessary covariance parameters. For all the datasets, minimizing the portfolio variance with the CV2 approach indeed leads to lower out-of-sample SD for the linear shrinkage methods LW\textsubscript{1} and LW\textsubscript{CC} and the GLASSO estimator. Especially noteworthy is the continuous and strong risk-reduction property of the GLASSO model alone and in combination with CV2. First, among the original models, the GLASSO estimation produces the best results for the 100SP and 250SP datasets. Second, when the sparsity parameter for the GLASSO estimation is selected with the CV2 approach, outperformance is superior for all the datasets. Hence, in respect to out-of-sample risk, applying graph models to induce sparsity within the precision matrix seems to be a valid approach, even in comparison to highly sophisticated methods such as the nonlinear shrinkage and approximate factor models. Interestingly, for the latter, the CV2 method does not lead to consistent outperformance in terms of risk. The original version of POET performs better than POET-CV2 in all cases but the 100SP dataset. In the case of GMV portfolios with short sales this result implies that the application of the CV with selection criteria such as SFE (CV1) and out-of-sample portfolio variance (CV2) does not result in a more optimal number of factors than the originally established function in Equation~\eqref{eq:covPOETK}. Moreover, for LW\textsubscript{NL}, there is almost no relevant difference in annualized out-of-sample SD. We can therefore argue that the efficiency of LW\textsubscript{NL} does not strongly depend on the choice of the kernel bandwidth's speed and hence, a data-driven specification cannot lead to an improvement in the performance out-of-sample.

For the CV1 approach, the investigation of the MSFE is mandatory. The values reported in Table~\ref{table:perf} indicate the distinct effect of the CV1 approach on the minimization of the MSFE out-of-sample. For all the estimation methods and datasets, except the isolated case of LW\textsubscript{CC} for 250SP, the MSFE is the lowest for the CV1 version of every estimator. Even robust estimators such as LW\textsubscript{NL} and POET exhibit higher forecasting power, measured by the MSFE, when the corresponding parameters are estimated with the CV1 approach. Nevertheless, it is noteworthy that the MSFE measure does not seem to proxy for the out-of-sample portfolio risk level. Within the financial literature including \citet{Zakamulin.2015}, the MSFE is studied in reference to datasets with low concentration ratios. Within a high-dimensional setting, however, a lower MSFE does not coincide with lower SD out-of-sample for any of the datasets or estimation methods.\footnote{Only the LW\textsubscript{NL} estimator for the 200SP and 250SP datasets yields the lowest risk levels and lowest MSFE for the CV1 approach. This result is merely based on a negligible difference and can thus be ignored.} Under the CV1 method, the SFE is computed as an estimator's squared distance to the monthly realized covariance matrix, calculated on the basis of daily returns (roughly 21 days) for $n$ assets. The implied concentration ratios, ranging from $50/21=2.38$ for the 50SP dataset to $250/21=11.90$ for the 250SP dataset, lead to ill-conditioned realized covariance matrices and a noisy SFE calculation.\footnote{To solve this problem, recent financial studies have focused on improving the estimation of large realized covariance matrices \citep[see, e.g.,][]{Hautsch.2012, Callot.2017, Bollerslev.2018}.} Therefore, we focus our further analysis on the CV2 approach.

To understand the magnitude of improvements in the CV2-based estimation methods as well as the superiority of the GLASSO method, Table~\ref{table:sign_250sp} presents the differences in annualized SDs and the respective pairwise significance levels across all the original covariance estimators and their CV2-based counterparts for the high-dimensional case of the 250SP dataset.\footnote{Appendix~\ref{app:gmv} compares further datasets. Overall, the results are similar in tendency, but are less pronounced due to a lower dimensionality in the data.} Table~\ref{table:sign_250sp} is to be read column-wise; that is, the difference in SD for the LW\textsubscript{1} and Sample estimator is listed under the second column for the first row. For completeness, we construct the table symmetrically. Still, we focus our attention on the elements above the diagonal only.

\begin{sidewaystable}[pt!]
	\centering
	\caption{Differences in SD p.a. of GMV-250SP across different estimators.\label{table:sign_250sp}}
	\resizebox{\linewidth}{!}{%
		\begin{threeparttable}
			\begin{tabular}{l *{11}{d{2.6}}}
				\toprule
				& \mc{Sample} &\mc{LW\textsubscript{1}} & \mc{LW\textsubscript{1}-CV2} & \mc{LW\textsubscript{CC}} & \mc{LW\textsubscript{CC}-CV2} & \mc{LW\textsubscript{NL}} & \mc{LW\textsubscript{NL}-CV2} & \mc{POET} & \mc{POET-CV2} & \mc{GLASSO} & \mc{GLASSO-CV2} \\
				\midrule
				Sample &  & \cellcolor[rgb]{0.045,0.815,0.193}-0.01420^{***} & \cellcolor[rgb]{0,0.8,0.2}-0.01967^{***} & \cellcolor[rgb]{0,0.8,0.2}-0.01557^{***} & \cellcolor[rgb]{0,0.8,0.2}-0.01903^{***} & \cellcolor[rgb]{0,0.8,0.2}-0.01955^{***} & \cellcolor[rgb]{0,0.8,0.2}-0.01948^{***} & \cellcolor[rgb]{0,0.8,0.2}-0.01746^{***} & \cellcolor[rgb]{0,0.8,0.2}-0.01715^{***} & \cellcolor[rgb]{0,0.8,0.2}-0.01988^{***} & \cellcolor[rgb]{0,0.8,0.2}-0.02075^{***} \\ 
				LW\textsubscript{1} & \cellcolor[rgb]{1,0.27,0}0.01420 &  & \cellcolor[rgb]{0.394,0.931,0.134}-0.00547^{***} & \cellcolor[rgb]{0.557,0.986,0.107}-0.00137^* & \cellcolor[rgb]{0.419,0.94,0.13}-0.00483^{***} & \cellcolor[rgb]{0.398,0.933,0.134}-0.00535^{***} & \cellcolor[rgb]{0.401,0.934,0.133}-0.00528^{***} & \cellcolor[rgb]{0.482,0.961,0.12}-0.00326^{***} & \cellcolor[rgb]{0.494,0.965,0.118}-0.00295^{***} & \cellcolor[rgb]{0.385,0.928,0.136}-0.00568^{***} & \cellcolor[rgb]{0.35,0.917,0.142}-0.00655^{***} \\ 
				LW\textsubscript{1}-CV2 & \cellcolor[rgb]{1,0.27,0}0.01967 & \cellcolor[rgb]{1,0.27,0}0.00547 &  & \cellcolor[rgb]{1,0.27,0}0.00410^{***} & \cellcolor[rgb]{0.861,1,0.279}0.00064 & \cellcolor[rgb]{0.811,1,0.378}0.00012 & \cellcolor[rgb]{0.818,1,0.365}0.00019 & \cellcolor[rgb]{1,0.979,0}0.00220^* & \cellcolor[rgb]{1,0.908,0}0.00252^{**} & \cellcolor[rgb]{0.658,1,0.187}-0.00021 & \cellcolor[rgb]{0.569,0.99,0.105}-0.00108 \\ 
				LW\textsubscript{CC} & \cellcolor[rgb]{1,0.27,0}0.01557 & \cellcolor[rgb]{0.93,1,0.141}0.00137 & \cellcolor[rgb]{0.448,0.949,0.125}-0.00410 &  & \cellcolor[rgb]{0.474,0.958,0.121}-0.00346^{***} & \cellcolor[rgb]{0.453,0.951,0.125}-0.00398^{***} & \cellcolor[rgb]{0.456,0.952,0.124}-0.00391^{***} & \cellcolor[rgb]{0.536,0.979,0.111}-0.00189^{**} & \cellcolor[rgb]{0.549,0.983,0.109}-0.00158^* & \cellcolor[rgb]{0.44,0.947,0.127}-0.00431^{**} & \cellcolor[rgb]{0.405,0.935,0.133}-0.00518^{***} \\ 
				LW\textsubscript{CC}-CV2 & \cellcolor[rgb]{1,0.27,0}0.01903 & \cellcolor[rgb]{1,0.27,0}0.00483 & \cellcolor[rgb]{0.586,0.995,0.102}-0.00064 & \cellcolor[rgb]{1,0.57,0}0.00346 &  & \cellcolor[rgb]{0.591,0.997,0.102}-0.00052 & \cellcolor[rgb]{0.594,0.998,0.101}-0.00045 & \cellcolor[rgb]{0.948,1,0.104}0.00156 & \cellcolor[rgb]{0.978,1,0.044}0.00188 & \cellcolor[rgb]{0.578,0.993,0.104}-0.00085 & \cellcolor[rgb]{0.543,0.981,0.11}-0.00173 \\ 
				LW\textsubscript{NL} & \cellcolor[rgb]{1,0.27,0}0.01955 & \cellcolor[rgb]{1,0.27,0}0.00535 & \cellcolor[rgb]{0.722,1,0.283}-0.00012 & \cellcolor[rgb]{1,0.309,0}0.00398 & \cellcolor[rgb]{0.85,1,0.301}0.00052 &  & \cellcolor[rgb]{0.807,1,0.387}0.00007 & \cellcolor[rgb]{0.998,1,0.005}0.00209^{**} & \cellcolor[rgb]{1,0.934,0}0.00240^{***} & \cellcolor[rgb]{0.599,1,0.1}-0.00033 & \cellcolor[rgb]{0.564,0.988,0.106}-0.00120 \\ 
				LW\textsubscript{NL}-CV2 & \cellcolor[rgb]{1,0.27,0}0.01948 & \cellcolor[rgb]{1,0.27,0}0.00528 & \cellcolor[rgb]{0.674,1,0.212}-0.00019 & \cellcolor[rgb]{1,0.344,0}0.00391 & \cellcolor[rgb]{0.843,1,0.314}0.00045 & \cellcolor[rgb]{0.753,1,0.329}-0.00007 &  & \cellcolor[rgb]{0.991,1,0.018}0.00202^{**} & \cellcolor[rgb]{1,0.95,0}0.00233^{***} & \cellcolor[rgb]{0.596,0.999,0.101}-0.00040 & \cellcolor[rgb]{0.561,0.987,0.106}-0.00127 \\ 
				POET & \cellcolor[rgb]{1,0.27,0}0.01746 & \cellcolor[rgb]{1,0.667,0}0.00326 & \cellcolor[rgb]{0.524,0.975,0.113}-0.00220 & \cellcolor[rgb]{0.979,1,0.041}0.00189 & \cellcolor[rgb]{0.549,0.983,0.108}-0.00156 & \cellcolor[rgb]{0.529,0.976,0.112}-0.00209 & \cellcolor[rgb]{0.531,0.977,0.111}-0.00202 &  & \cellcolor[rgb]{0.83,1,0.34}0.00032 & \cellcolor[rgb]{0.515,0.972,0.114}-0.00242 & \cellcolor[rgb]{0.481,0.96,0.12}-0.00329^{**} \\ 
				POET-CV2 & \cellcolor[rgb]{1,0.27,0}0.01715 & \cellcolor[rgb]{1,0.811,0}0.00295 & \cellcolor[rgb]{0.511,0.97,0.115}-0.00252 & \cellcolor[rgb]{0.949,1,0.101}0.00158 & \cellcolor[rgb]{0.537,0.979,0.111}-0.00188 & \cellcolor[rgb]{0.516,0.972,0.114}-0.00240 & \cellcolor[rgb]{0.519,0.973,0.114}-0.00233 & \cellcolor[rgb]{0.599,1,0.1}-0.00032 &  & \cellcolor[rgb]{0.503,0.968,0.116}-0.00273^* & \cellcolor[rgb]{0.468,0.956,0.122}-0.00361^{**} \\ 
				GLASSO & \cellcolor[rgb]{1,0.27,0}0.01988 & \cellcolor[rgb]{1,0.27,0}0.00568 & \cellcolor[rgb]{0.82,1,0.36}0.00021 & \cellcolor[rgb]{1,0.27,0}0.00431 & \cellcolor[rgb]{0.881,1,0.239}0.00085 & \cellcolor[rgb]{0.831,1,0.338}0.00033 & \cellcolor[rgb]{0.838,1,0.325}0.00040 & \cellcolor[rgb]{1,0.931,0}0.00242 & \cellcolor[rgb]{1,0.86,0}0.00273 &  & \cellcolor[rgb]{0.577,0.992,0.104}-0.00087^{**} \\ 
				GLASSO-CV2 & \cellcolor[rgb]{1,0.27,0}0.02075 & \cellcolor[rgb]{1,0.27,0}0.00655 & \cellcolor[rgb]{0.903,1,0.195}0.00108 & \cellcolor[rgb]{1,0.27,0}0.00518 & \cellcolor[rgb]{0.963,1,0.073}0.00173 & \cellcolor[rgb]{0.914,1,0.172}0.00120 & \cellcolor[rgb]{0.92,1,0.159}0.00127 & \cellcolor[rgb]{1,0.654,0}0.00329 & \cellcolor[rgb]{1,0.497,0}0.00361 & \cellcolor[rgb]{0.883,1,0.235}0.00087 &  \\  \midrule
				better than \% of models & 0.0 & 0.1 & 0.8 & 0.2 & 0.5 & 0.7 & 0.6 & 0.4 & 0.3 & 0.9 & 1.0 \\ 
				\bottomrule
			\end{tabular}
			\begin{tablenotes}
				\item This table shows the differences in the annualized out-of-sample SD of the GMV-250SP portfolios across the main covariance estimation methods and their CV2-based counterparts. The table is constructed in a symmetrical way with an applied color scheme from red (higher SD than the other model) to green (lower SD than the other model). In addition, on the elements above the diagonal, the significant pairwise outperformance in terms of variance is denoted by asterisks: *** denotes significance at the 0.001 level; ** denotes significance at the 0.01 level; and * denotes significance at the 0.05 level. Finally, for each model, we report the percentage of the other models that exhibit higher variance as a qualitative measure.
			\end{tablenotes}
		\end{threeparttable}
	}
	\vspace{1cm} 
	\caption{Differences in SD p.a. of GMV-NOSHORT-250SP across different estimators\label{table:sign_250sp_noshort}}
	\resizebox{\linewidth}{!}{%
		\begin{threeparttable}
			\begin{tabular}{l *{11}{d{2.6}}}
				\toprule
				& \mc{Sample} &\mc{LW\textsubscript{1}} & \mc{LW\textsubscript{1}-CV2} & \mc{LW\textsubscript{CC}} & \mc{LW\textsubscript{CC}-CV2} & \mc{LW\textsubscript{NL}} & \mc{LW\textsubscript{NL}-CV2} & \mc{POET} & \mc{POET-CV2} & \mc{GLASSO} & \mc{GLASSO-CV2} \\
				\midrule
				Sample &  & \cellcolor[rgb]{0.85,1,0.299}-0.00001 & \cellcolor[rgb]{0.729,1,0.068}-0.00016 & \cellcolor[rgb]{0.134,0.845,0.178}-0.00051 & \cellcolor[rgb]{0.829,0.971,0.341}0.00005 & \cellcolor[rgb]{1,0.27,0}0.00051 & \cellcolor[rgb]{1,0.392,0}0.00047 & \cellcolor[rgb]{0.307,0.902,0.149}-0.00040^{***} & \cellcolor[rgb]{0.191,0.864,0.168}-0.00047^{***} & \cellcolor[rgb]{0.935,0.865,0.129}0.00023 & \cellcolor[rgb]{0,0.8,0.2}-0.00060 \\ 
				LW\textsubscript{1} & \cellcolor[rgb]{0.805,0.995,0.39}0.00001 &  & \cellcolor[rgb]{0.746,1,0.064}-0.00015 & \cellcolor[rgb]{0.148,0.849,0.175}-0.00050 & \cellcolor[rgb]{0.834,0.966,0.331}0.00006 & \cellcolor[rgb]{1,0.27,0}0.00052^* & \cellcolor[rgb]{1,0.367,0}0.00048^* & \cellcolor[rgb]{0.321,0.907,0.147}-0.00039 & \cellcolor[rgb]{0.204,0.868,0.166}-0.00046^* & \cellcolor[rgb]{0.94,0.86,0.119}0.00024 & \cellcolor[rgb]{0,0.8,0.2}-0.00059 \\ 
				LW\textsubscript{1}-CV2 & \cellcolor[rgb]{0.896,0.904,0.207}0.00016 & \cellcolor[rgb]{0.891,0.909,0.217}0.00015 &  & \cellcolor[rgb]{0.401,0.934,0.133}-0.00034 & \cellcolor[rgb]{0.926,0.874,0.149}0.00021 & \cellcolor[rgb]{1,0.27,0}0.00067^* & \cellcolor[rgb]{1,0.27,0}0.00063 & \cellcolor[rgb]{0.575,0.992,0.104}-0.00024 & \cellcolor[rgb]{0.458,0.953,0.124}-0.00031^* & \cellcolor[rgb]{1,0.636,0}0.00039 & \cellcolor[rgb]{0.249,0.883,0.159}-0.00044 \\ 
				LW\textsubscript{CC} & \cellcolor[rgb]{1,0.28,0}0.00051 & \cellcolor[rgb]{1,0.306,0}0.00050 & \cellcolor[rgb]{1,0.779,0}0.00034 &  & \cellcolor[rgb]{1,0.27,0}0.00055 & \cellcolor[rgb]{1,0.27,0}0.00101 & \cellcolor[rgb]{1,0.27,0}0.00097 & \cellcolor[rgb]{0.862,0.938,0.275}0.00010 & \cellcolor[rgb]{0.82,0.98,0.359}0.00003 & \cellcolor[rgb]{1,0.27,0}0.00073 & \cellcolor[rgb]{0.875,1,0.031}-0.00009 \\ 
				LW\textsubscript{CC}-CV2 & \cellcolor[rgb]{0.965,1,0.009}-0.00005 & \cellcolor[rgb]{0.948,1,0.013}-0.00006 & \cellcolor[rgb]{0.625,1,0.094}-0.00021 & \cellcolor[rgb]{0.052,0.817,0.191}-0.00055 &  & \cellcolor[rgb]{1,0.42,0}0.00046 & \cellcolor[rgb]{1,0.544,0}0.00042 & \cellcolor[rgb]{0.226,0.875,0.162}-0.00045 & \cellcolor[rgb]{0.109,0.836,0.182}-0.00052 & \cellcolor[rgb]{0.906,0.894,0.188}0.00018 & \cellcolor[rgb]{0,0.8,0.2}-0.00065 \\ 
				LW\textsubscript{NL} & \cellcolor[rgb]{0.127,0.842,0.179}-0.00051 & \cellcolor[rgb]{0.114,0.838,0.181}-0.00052 & \cellcolor[rgb]{0,0.8,0.2}-0.00067 & \cellcolor[rgb]{0,0.8,0.2}-0.00101 & \cellcolor[rgb]{0.209,0.87,0.165}-0.00046 &  & \cellcolor[rgb]{0.984,1,0.004}-0.00004^{**} & \cellcolor[rgb]{0,0.8,0.2}-0.00091^* & \cellcolor[rgb]{0,0.8,0.2}-0.00098^{**} & \cellcolor[rgb]{0.503,0.968,0.116}-0.00028 & \cellcolor[rgb]{0,0.8,0.2}-0.00111 \\ 
				LW\textsubscript{NL}-CV2 & \cellcolor[rgb]{0.194,0.865,0.168}-0.00047 & \cellcolor[rgb]{0.181,0.86,0.17}-0.00048 & \cellcolor[rgb]{0,0.8,0.2}-0.00063 & \cellcolor[rgb]{0,0.8,0.2}-0.00097 & \cellcolor[rgb]{0.276,0.892,0.154}-0.00042 & \cellcolor[rgb]{0.824,0.976,0.352}0.00004 &  & \cellcolor[rgb]{0,0.8,0.2}-0.00087^* & \cellcolor[rgb]{0,0.8,0.2}-0.00094^* & \cellcolor[rgb]{0.57,0.99,0.105}-0.00024 & \cellcolor[rgb]{0,0.8,0.2}-0.00107 \\ 
				POET & \cellcolor[rgb]{1,0.603,0}0.00040 & \cellcolor[rgb]{1,0.628,0}0.00039 & \cellcolor[rgb]{0.942,0.858,0.116}0.00024 & \cellcolor[rgb]{0.849,1,0.038}-0.00010 & \cellcolor[rgb]{1,0.451,0}0.00045 & \cellcolor[rgb]{1,0.27,0}0.00091 & \cellcolor[rgb]{1,0.27,0}0.00087 &  & \cellcolor[rgb]{0.921,1,0.02}-0.00007 & \cellcolor[rgb]{1,0.27,0}0.00063 & \cellcolor[rgb]{0.654,1,0.087}-0.00020 \\ 
				POET-CV2 & \cellcolor[rgb]{1,0.385,0}0.00047 & \cellcolor[rgb]{1,0.411,0}0.00046 & \cellcolor[rgb]{0.984,0.816,0.032}0.00031 & \cellcolor[rgb]{0.997,1,0.001}-0.00003 & \cellcolor[rgb]{1,0.27,0}0.00052 & \cellcolor[rgb]{1,0.27,0}0.00098 & \cellcolor[rgb]{1,0.27,0}0.00094 & \cellcolor[rgb]{0.842,0.958,0.316}0.00007 &  & \cellcolor[rgb]{1,0.27,0}0.00070 & \cellcolor[rgb]{0.803,1,0.049}-0.00013 \\ 
				GLASSO & \cellcolor[rgb]{0.592,0.997,0.101}-0.00023 & \cellcolor[rgb]{0.579,0.993,0.104}-0.00024 & \cellcolor[rgb]{0.325,0.908,0.146}-0.00039 & \cellcolor[rgb]{0,0.8,0.2}-0.00073 & \cellcolor[rgb]{0.694,1,0.076}-0.00018 & \cellcolor[rgb]{0.967,0.833,0.065}0.00028 & \cellcolor[rgb]{0.943,0.857,0.113}0.00024 & \cellcolor[rgb]{0,0.8,0.2}-0.00063 & \cellcolor[rgb]{0,0.8,0.2}-0.00070 &  & \cellcolor[rgb]{0,0.8,0.2}-0.00083^{***} \\ 
				GLASSO-CV2 & \cellcolor[rgb]{1,0.27,0}0.00060 & \cellcolor[rgb]{1,0.27,0}0.00059 & \cellcolor[rgb]{1,0.494,0}0.00044 & \cellcolor[rgb]{0.855,0.945,0.29}0.00009 & \cellcolor[rgb]{1,0.27,0}0.00065 & \cellcolor[rgb]{1,0.27,0}0.00111 & \cellcolor[rgb]{1,0.27,0}0.00107 & \cellcolor[rgb]{0.917,0.883,0.165}0.00020 & \cellcolor[rgb]{0.875,0.925,0.249}0.00013 & \cellcolor[rgb]{1,0.27,0}0.00083 &  \\ \midrule
				better than \% of models & 0.4 & 0.5 & 0.6 & 0.9 & 0.3 & 0.0 & 0.1 & 0.7 & 0.8 & 0.2 & 1.0 \\ 
				\bottomrule
			\end{tabular}
			\begin{tablenotes}
				\item This table shows the differences in the annualized out-of-sample SD of the GMV-NOSHORT-250SP portfolios across the main covariance estimation methods and their CV2-based counterparts. The table is constructed in a symmetrical way with an applied color scheme from red (higher SD than the other model) to green (lower SD than the other model). In addition, on the elements above the diagonal, the significant pairwise outperformance in terms of variance is denoted by asterisks: *** denotes significance at the 0.001 level; ** denotes significance at the 0.01 level; and * denotes significance at the 0.05 level. Finally, for each model, we report the percentage of the other models that exhibit higher variance as a qualitative measure.   
			\end{tablenotes}
		\end{threeparttable}
	}
\end{sidewaystable}

At first glance, we can once more distinguish the weak performance of the sample covariance estimator. All the other estimators lead to significantly less out-of-sample risk with a p-value of roughly 0.001 or lower. The second worst estimation method for this asset universe is the original LW\textsubscript{1}, followed by LW\textsubscript{CC}. On the contrary, when the linear shrinkage intensity is optimized for with respect to the out-of-sample portfolio variance with CV2, we observe an astonishing improvement for those estimators. Both LW\textsubscript{1}-CV2 and LW\textsubscript{CC}-CV2 result in a significantly lower out-of-sample SD than their original counterparts. Comparing LW\textsubscript{CC}-CV2 with LW\textsubscript{1}-CV2, we can conclude that linear shrinkage toward the identity matrix yields more stable and efficient portfolio allocation than the same technique applied to a constant correlation model. It seems, therefore, that within the data-driven approach it is better to assume less than assume the wrong structure. Another surprising insight emerges from the comparison of LW\textsubscript{1}-CV2 with LW\textsubscript{NL}. Although especially designed to overcome the high-dimensionality problem, both the original and CV2-based nonlinear shrinkage methods lead to higher out-of-sample risk than the data-driven linear shrinkage estimators. This effect is observable for the 100SP dataset as well (see, for reference, Table~\ref{table:sign_100sp}). As the difference is not statistically significant in any of the cases, we can only draw a qualitative conclusion that a methodologically easy-to-understand and simple-to-implement method can perform as well as a complex state-of-the-art estimator when the necessary tuning parameter (here, the shrinkage intensity) is identified in a data-driven way. With respect to the GLASSO estimation method, we find that GLASSO with CV2 results in a significantly lower out-of-sample SD than the original estimator (p-value of 0.01). This finding reveals that a suitable selection criterion for the sparsity parameter is of utmost importance.	Furthermore, GLASSO-CV2 yields significantly lower out-of-sample portfolio variance than the sample, LW\textsubscript{1}, LW\textsubscript{CC}, and POET estimators. LW\textsubscript{NL} is significantly outperformed by GLASSO-CV2 for the 50SP and 100SP datasets with a p-value of 0.001 (see, for reference, the significant levels in Table~\ref{table:sign_50sp} and Table~\ref{table:sign_100sp}). Overall, these results confirm that GLASSO in combination with CV2 is the most efficient covariance estimator among the estimators considered in our study.

Table~\ref{table:perf} additionally reports the average monthly turnover rate as a proxy for the arising transaction costs induced by monthly rebalancing. The Naive portfolio, being long only and equally-weighted by construction, naturally has the lowest turnover (approximately 0.06 on average across all the datasets). As expected, the GMV portfolios estimated with the sample covariance matrix are characterized by extreme exposures for all the datasets. With higher dimensionality the ill-conditioned sample estimator induces even stronger dispersion in the portfolio weights. On the other hand, an estimation with GLASSO or its CV-based equivalents has the most pronounced positive effect on the allocation characteristics of the GMV portfolio. In particular, the GLASSO-CV1 estimation methodology results in GMV portfolios with the lowest turnover for the 200SP and 250SP datasets. For the least high-dimensional dataset, 50SP, GLASSO-CV2 leads to the lowest turnover. An interesting case is the 100SP dataset, where clear outperformance in terms of turnover rate occurs for the estimator LW\textsubscript{1}-CV2, followed closely by GLASSO-CV1. It seems that when the concentration ratio is tolerable, as for the case of 100SP, the linear shrinkage methodology, as a convex combination between the sample covariance and an identity matrix, produces satisfactory results. This can be explained by the fact that the underlying model in LW\textsubscript{1} is equivalent to the introduction of a ridge type penalty in the estimation \citep{Warton.2008}, which has been proven to induce stability. However, when the sample covariance matrix becomes ill-conditioned, as for the 200SP and 250SP datasets, even a sophisticated data-driven choice of the linear shrinkage intensity cannot outperform the GLASSO estimation in terms of turnover. While LW\textsubscript{1} shrinks the sample covariance matrix toward the identity matrix, GLASSO shrinks the precision matrix toward the identity matrix. Since the Naive portfolio corresponds to a GMV portfolio estimated with an identity covariance and hence, precision matrix, one may suggest that both estimation methods result in an implicit shrinkage of the sample GMV portfolio weights toward an equally-weighted portfolio, as in \citet{Tu.2011}, and therefore perform well in terms of turnover. Surprisingly, the estimator LW\textsubscript{NL} is strongly outperformed by all the data-driven estimators except POET with CV1 and CV2. For example, while GLASSO achieves a turnover rate of 0.61 for 250SP, estimation with LW\textsubscript{NL} leads to 0.73 average monthly turnover. The application of CV1 and CV2 for determining the sparsity parameter within the GLASSO model amplifies this result. More importantly, when a covariance estimator is susceptible to an improvement by the data-driven estimation of the necessary parameters, as LW\textsubscript{1}, LW\textsubscript{CC}, and GLASSO are, the implementation of CV leads to a strong positive impact on the stability of optimal weights.

\subsection{GMV without short sales}\label{subsec:resultsgmvnoshort}

Focusing on more practically relevant portfolio strategies, we construct a second set of GMV portfolios that do not exhibit negative weights (GMV-NOSHORT). The exclusion of short sales is a common regulatory constraint that strongly influences the optimal performance in respect to the out-of-sample risk and the allocation of weights. To investigate those, Table~\ref{table:perf_noshort} reports the main out-of-sample measures for GMV-NOSHORT. Since the examined short-sale constraint does not play any role in the CV1-based estimation of the covariance matrix, we do not report the MSFE values. The table is structured similarly to Table~\ref{table:perf} with the columns representing the investment universes (50SP, 100SP, 200SP and 250SP), and performance measures, while the rows indicate the portfolio strategies based on the considered covariance estimation methods.

\begin{table}
	\centering
	\caption{Performance of GMV-NOSHORT portfolios across different estimators and datasets.\label{table:perf_noshort}}
	\resizebox{\textwidth}{!}{%
		\begin{threeparttable}
			\begin{tabular}{l*{2}c@{\hskip 0.3in}*2{c}@{\hskip 0.3in}*2{c}@{\hskip 0.3in}*2{c}}
				\toprule
				\multicolumn{2}{r@{\hskip -0.15in}}{50SP}  & \multicolumn{2}{r@{\hskip -0.15in}}{100SP} & \multicolumn{2}{r@{\hskip -0.15in}}{200SP} & \multicolumn{2}{r@{\hskip -0.15in}}{250SP}  \\
				\midrule
				& SD & TO & SD & TO & SD & TO & SD & TO \\ \midrule
				Naive & 0.1887 & 0.0596  & 0.1811 & 0.0587 & 0.1852 & 0.0601 & 0.1858 & 0.0592 \\ 
				Sample & 0.1295 & 0.1473 & 0.1168 & 0.1810 & 0.1137 &  0.2158 & 0.1105 & 0.2227 \\
				& & & & & & & & \\
				LW\textsubscript{1} & 0.1296 & \textbf{0.1382} & \textbf{0.1167} & \textbf{0.1708} & \textbf{0.1135} & \textbf{0.1974} & 0.1105 & \textbf{0.2055} \\ 
				LW\textsubscript{1}-CV1 & 0.1308 & 0.1486 &  0.1180 & 0.1712 & 0.1151 & 0.1999 & 0.1122 & 0.2310 \\ 
				LW\textsubscript{1}-CV2 & \textbf{0.1294} & 0.1578 & 0.1169 & 0.1896 & 0.1136 & 0.2287 & \textbf{0.1103} & 0.2640 \\
				& & & & & & & & \\
				LW\textsubscript{CC} & 0.1291 & 0.1347 & \textbf{0.1167} & 0.1655 & \textbf{0.1134} & 0.1941 & \textbf{0.1100} & \textbf{0.2011} \\ 
				LW\textsubscript{CC}-CV1 & 0.1297 & \underline{\textbf{0.1310}} & 0.1179 & \textbf{0.1607} & 0.1145 & \textbf{0.1863} & 0.1110 & 0.2246 \\ 
				LW\textsubscript{CC}-CV2 & \textbf{0.1286} & 0.1484 & 0.1169 & 0.1857 & 0.1134 & 0.2228 & 0.1105 & 0.2749 \\ 
				& & & & & & & & \\
				LW\textsubscript{NL} & 0.1296 & \textbf{0.1351} & 0.1166 & \underline{\textbf{0.1589}} & 0.1136 & \textbf{0.1770} & 0.1110 & \textbf{0.1806} \\ 
				LW\textsubscript{NL}-CV1  & 0.1296 & 0.1360 & 0.1166 & 0.1602 & 0.1136 & 0.1779 & 0.1110 & 0.1815 \\ 
				LW\textsubscript{NL}-CV2 & \textbf{0.1295} & 0.1361 & \textbf{0.1166} & 0.1600 & \textbf{0.1135} & 0.1776 & \textbf{0.1110} & 0.1817 \\ 
				& & & & & & & & \\
				POET & 0.1291 & \textbf{0.1392} & 0.1165 & \textbf{0.1711} & 0.1135 & \textbf{0.2034} & 0.1101 & \textbf{0.2125} \\ 
				POET-CV1  & \textbf{0.1290} & 0.1418 & 0.1164 & 0.1752 & 0.1134 & 0.2038 & 0.1101 & 0.2167 \\ 
				POET-CV2 & 0.1291 & 0.1452 & \textbf{0.1164} & 0.1759 & \textbf{0.1134} & 0.2069 & \textbf{0.1100} & 0.2283 \\ 
				& & & & & & & & \\
				GLASSO & 0.1302 & 0.2828 & 0.1165 & 0.2622 & 0.1134 & 0.2368 & 0.1107 & 0.2704 \\ 
				GLASSO-CV1 & 0.1314 & 0.1468 & 0.1196 & 0.1752 & 0.1182 & 0.1934 & 0.1158 & 0.2000 \\ 
				GLASSO-CV2 & \underline{\textbf{0.1285}} & \textbf{0.1348} & \underline{\textbf{0.1158}} & \textbf{0.1693} & \underline{\textbf{0.1132}} & \underline{\textbf{0.1690}} & \underline{\textbf{0.1099}} & \underline{\textbf{0.1778}} \\ 
				\bottomrule
			\end{tabular}
			\begin{tablenotes}
				\item This table reports the annualized out-of-sample SD and average monthly turnover rate (TO) of the GMV-NOSHORT portfolios across all the considered datasets with 50, 100, 200 and 250 stocks, respectively. We report the lowest SD and TO for each estimation method in bold. The best results in terms of SD for each dataset are underlined. We additionally underline the lowest TO, excluding the Naive portfolio. 
			\end{tablenotes}
		\end{threeparttable}
	}
\end{table}

First, the sample covariance estimator leads to better results in comparison to Table~\ref{table:perf}, verifying the impact of constraints on the minimization of estimation errors in portfolio weights, as shown by \citet{Jagannathan.2003}. In contrast to the previous results, the differences in out-of-sample performance in terms of portfolio risk are generally less distinctive among the estimation methods. While GLASSO-CV2 continues to achieve the lowest out-of-sample annualized SD for all the datasets, a data-driven approach only inconsistently enhances the performance of the other original estimation methods. For instance, both LW\textsubscript{1} and LW\textsubscript{CC} perform better than their CV2-based counterparts for the 100SP and 200SP datasets. Although both the nonlinear shrinkage and POET with CV2-estimated parameters seem to produce lower out-of-sample risk, the effect is negligible.\footnote{The relative differences in SD among all the original and cross-validated estimators are presented in Figure~\ref{fig:sdcomparison_noshort}.} Moreover, Table~\ref{table:sign_250sp_noshort} presents the differences in annualized SDs and the respective pairwise significance levels across all the main covariance estimation methods and their CV2-based counterparts for the high-dimensional case of the 250SP dataset.\footnote{Appendix~\ref{app:gmv_noshort} compares further datasets.} The table is constructed similarly to Table~\ref{table:sign_250sp}.

The first notable consequence of the short-sale constraint is the improvement in portfolio performance for the case of the sample covariance matrix. Table~\ref{table:sign_250sp_noshort} shows that only the POET estimators with and without CV2 yield a statistically lower out-of-sample SD than the sample covariance matrix with a p-value of 0.001.	Both the original and the CV2-based POET estimation methods result in a lower out-of-sample portfolio variance than all the other methods except LW\textsubscript{CC} and GLASSO-CV2. However, although GLASSO-CV2 outperforms all the other estimators qualitatively, the results are significant only compared to the original GLASSO estimator with a p-value of 0.001. This once more confirms that, first, the sparsity parameter for GLASSO is best determined with a CV and, second, the selection criterion within the CV should match the performance measure. In general, when no short sales are allowed, POET-CV2 and GLASSO-CV2 consistently outperform the rest of the methods (see, for reference, Tables~\ref{table:sign_100sp_noshort} and \ref{table:sign_200sp_noshort} for the other datasets). For a less high-dimensional dataset, such as 100SP, the out-of-sample SD generated with GLASSO-CV2 is even significantly lower than the out-of-sample SD induced by POET. Although less pronounced, the positive effects of a CV methodology with an adequate selection criterion, namely the variance of a short sales constrained GMV, are noticeable here as well. In particular, the usage of GLASSO-CV2 significantly improves the out-of-sample risk of a GMV without short sales for the 50SP, 100SP, and 250SP datasets.\footnote{For 200SP, GLASSO-CV2 outperforms all the other methods in terms of SD; however, the results are not significant.}

Finally, we examine the average monthly turnover rates, reported in Table~\ref{table:perf_noshort}. Since the portfolios are optimized with the additional no short-sale constraint, we expect the optimal weights to be much less dispersed across the rebalancing periods. Such behavior in turn results in lower turnover rates and transaction costs. In comparison to the GMV portfolios from Subsection~\ref{subsec:resultsgmv}, the turnover reduction is present even for the case of the ill-conditioned sample covariance estimator. The latter still performs worst, compared to the other methods with more dispersion in the weights for high-dimensional datasets such as 200SP and 250SP. In addition, for those datasets, we observe the lowest turnover rates when GLASSO-CV2 is used, followed by LW\textsubscript{NL} with a difference of one percent. We therefore continue to observe the positive effects of a data-driven estimation with the portfolio variance as a measure of fit not only on the out-of-sample risk but as well on the turnover rates.

\section{Conclusion}\label{sec:conclusion}

In this study, we review some of the most recent and efficient estimation methods for high-dimensional minimum-variance portfolios.
We extend the current research by proposing a data-driven methodology to determine the corresponding tuning parameters such as the linear shrinkage intensity and the sparsity penalty term. 

In a detailed empirical analysis with four datasets, we identify the characteristics of our data-driven methodology. First, we establish that the selection criterion within the CV should correspond to the performance measure of interest. We show that the lowest overall out-of-sample portfolio risk is indeed generated when we select the optimal tuning parameters by minimizing the portfolio variance with the proposed CV.
In particular, the application of this procedure to each of the considered estimators with the exception of the nonlinear shrinkage leads to superior GMV or GMV-NOSHORT portfolios in terms of out-of-sample SD and average turnover rates. The performance of the nonlinear shrinkage estimator is only slightly affected by the speed of the kernel bandwidth and a data-driven selection of this speed does not lead to a significantly lower risk. Moreover, the POET estimator in combination with a CV technique seems to perform generally better under the scenario of short sales constraints. On the other hand, the GLASSO estimator clearly outperforms the other high-dimensional estimation methods for all data scenarios considered when calibrated accordingly. This not only provides new insights into the application of GLASSO in the portfolio context, but also confirms the relationship between lasso regulation and the power of CV in a covariance estimation context. We additionally demonstrate that a data-driven methodology is beneficial to estimators whose performance depends strongly on the embedded tuning parameters, as is the case with linear shrinkage and GLASSO estimation methods. Even complex and highly efficient estimators can be surpassed by simpler approaches if a sophisticated data-driven technique is used. One of the reasons for this observation is the rapid adaptation of the CV toward ever-changing market situations.

Within our analysis we investigate only high-dimensional covariance estimation methods that assume homoscedasticity in the returns.
Since we observe a time-variable parameter selection with the CV approach and a resulting improvement in the  out-of-sample performance, we argue that the combination of data-driven parameter evaluation and time-dependent high-dimensional variance estimators, as recently proposed by \citet{Halbleib.2014} and \citet{Engle.2019}, is an important topic for future research.

\begin{acknowledgements}
This is a pre-print of the article submitted to Financial Markets and Portfolio Management. You can find the accepted version online at: 10.1007/s11408-020-00370-4.
\end{acknowledgements}

\bibliographystyle{spbasic}      
\bibliography{CovMatrix}   

\clearpage
\appendix

\section{Covariance Parameters}\label{app:covparam}

\begin{figure}[h]
	\begin{center}
		\begin{minipage}{\textwidth}
			\subfigure[LW\textsubscript{CC}]{
				\resizebox*{.5\textwidth}{!}{\includegraphics[width=.99\linewidth]{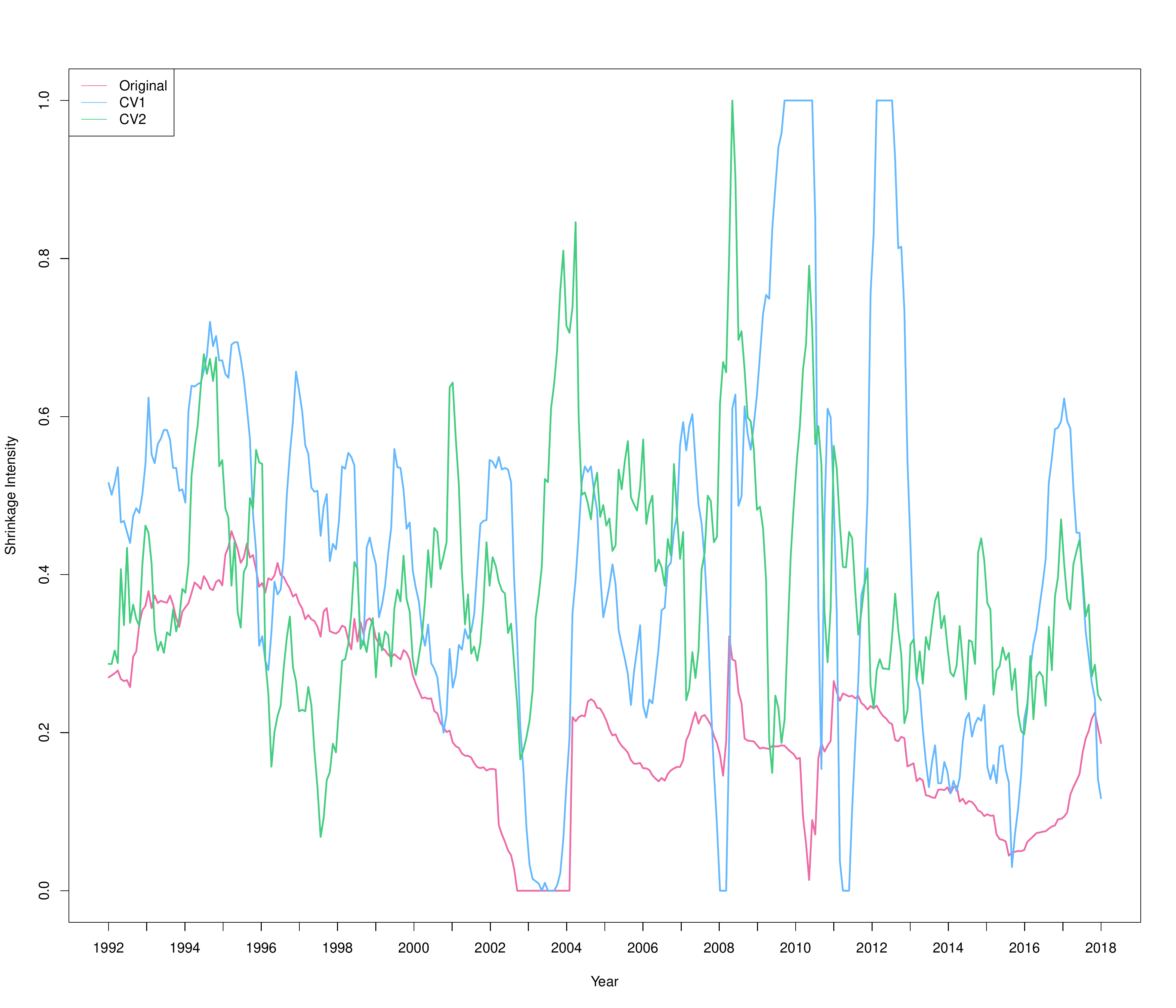}}\label{subfig:cc}}
			\subfigure[LW\textsubscript{NL}]{
				\resizebox*{.5\textwidth}{!}{\includegraphics[width=.99\linewidth]{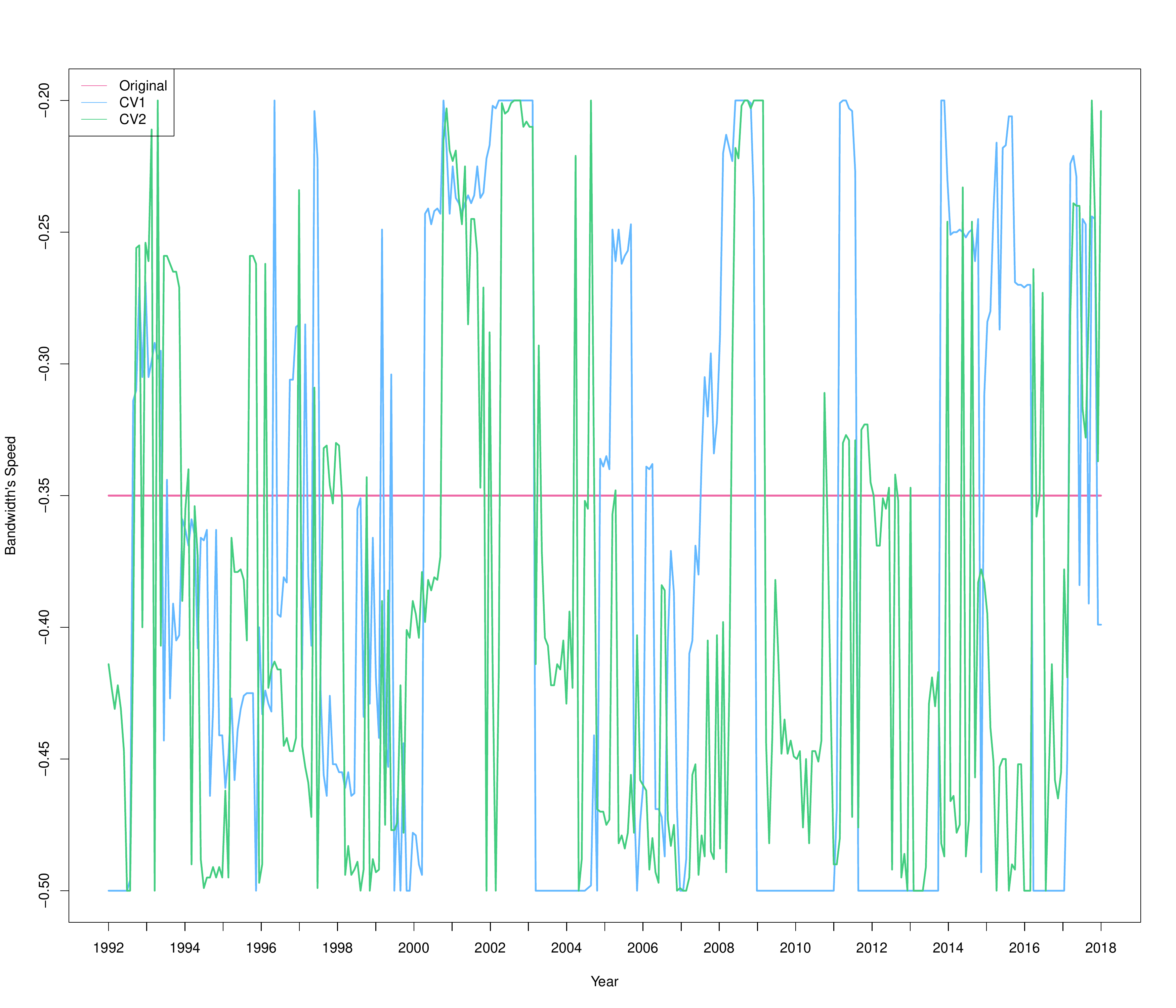}}\label{subfig:lwnl}}
		\end{minipage}
		\begin{minipage}{\textwidth}
			\subfigure[POET]{
				\resizebox*{.5\textwidth}{!}{\includegraphics[width=.99\linewidth]{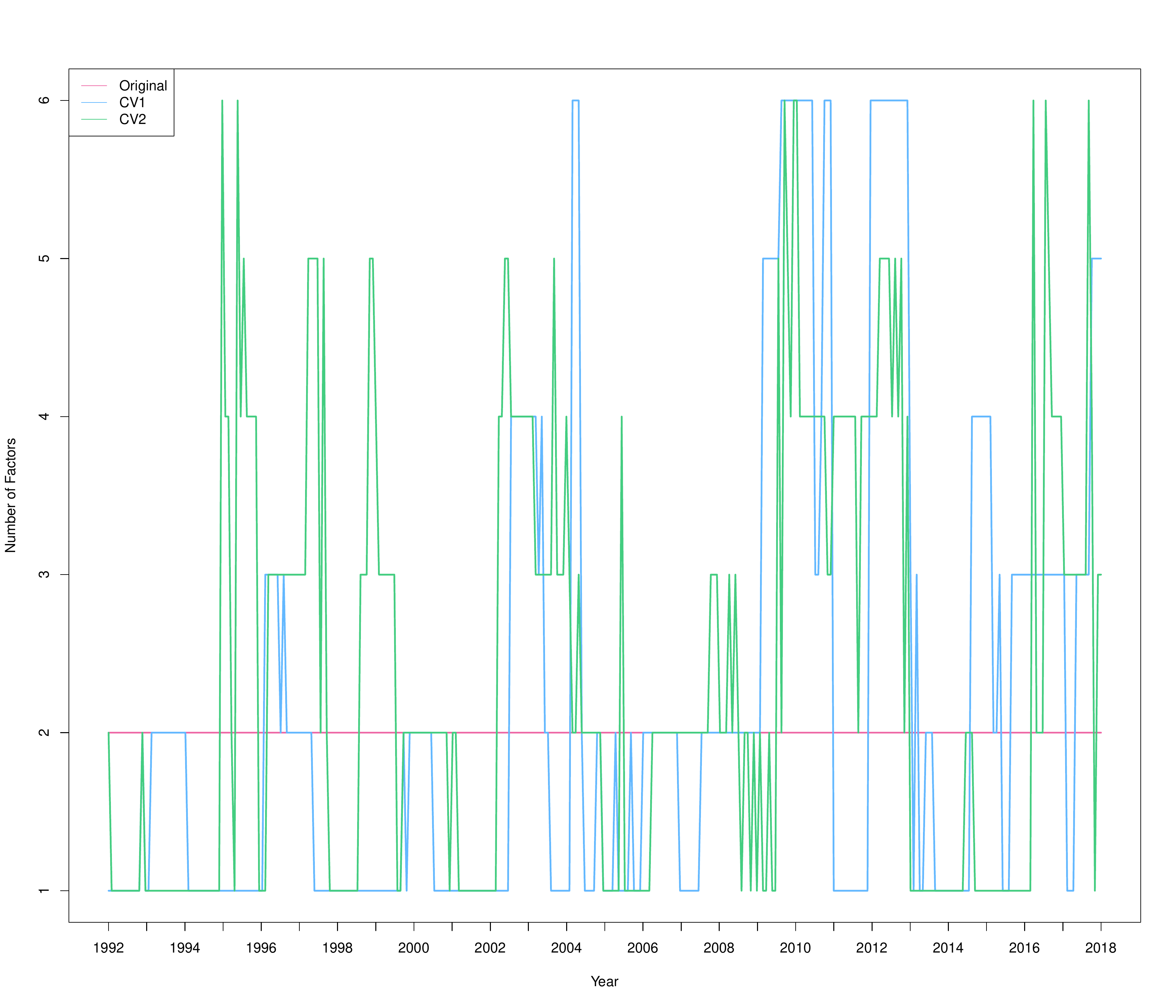}}\label{subfig:poet}}
			\subfigure[GLASSO]{
				\resizebox*{.5\textwidth}{!}{\includegraphics[width=.99\linewidth]{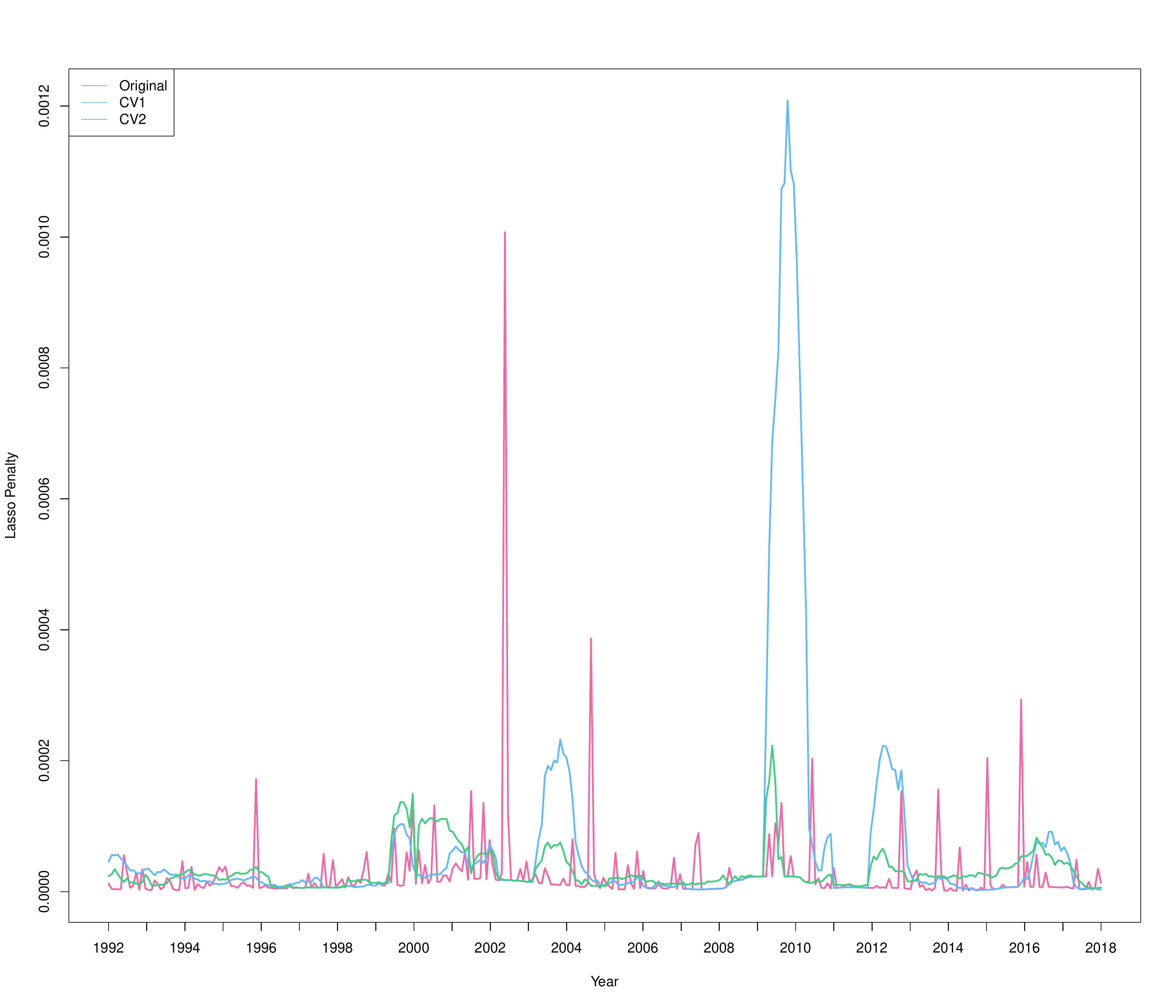}}\label{subfig:glasso}}
		\end{minipage}
		\caption{Optimally selected parameters with original, CV1, and CV2 covariance estimation methods for the 50SP dataset.}\label{fig:covparameters}
	\end{center}
\end{figure}

\clearpage
\section{GMV with short sales}\label{app:gmv}

\begin{figure}[h]
	\begin{center}
		\resizebox*{\textwidth}{!}{\includegraphics{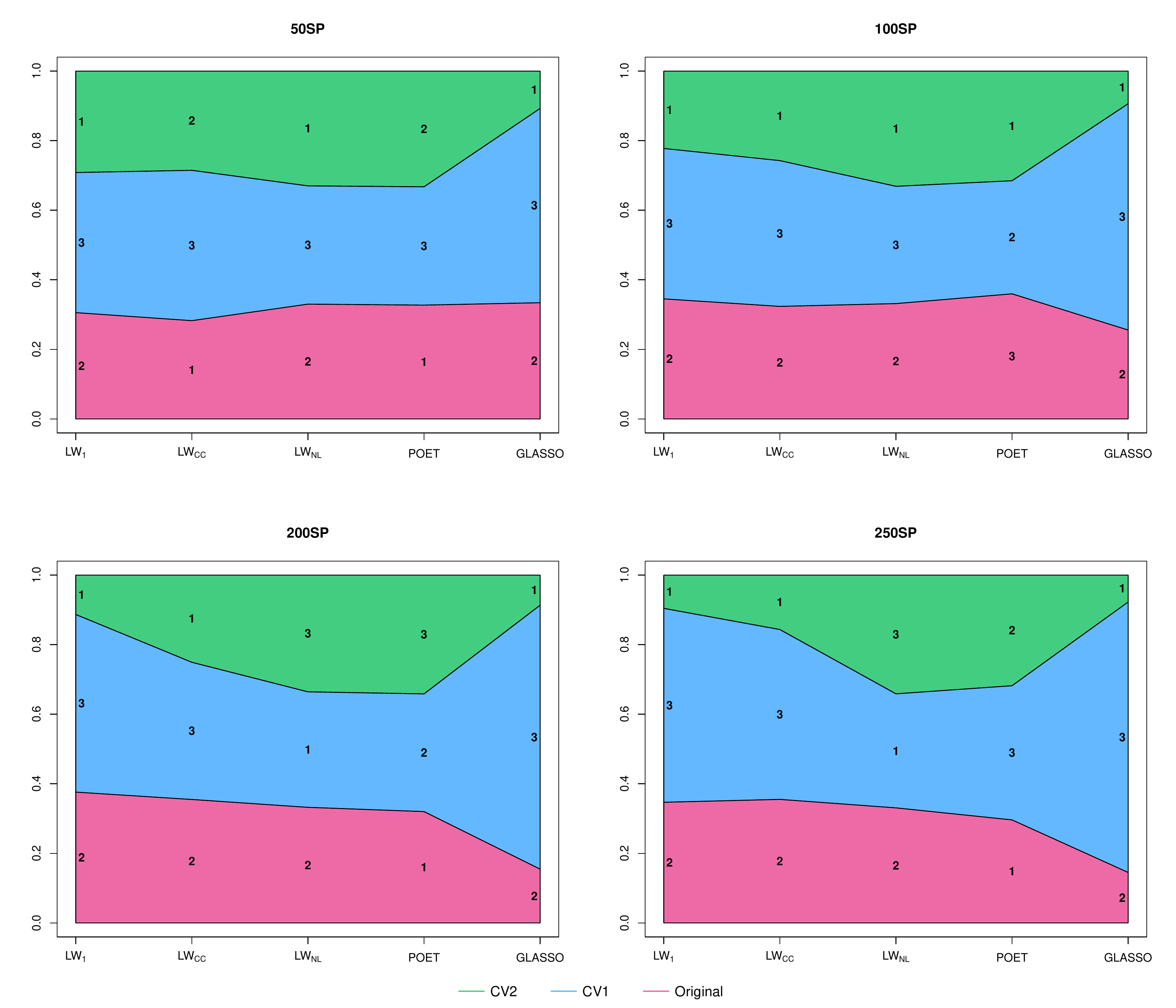}}
		\caption{Relative differences in the annualized SD of the GMV portfolios across all the covariance estimators.}\label{fig:sdcomparison}
	\end{center}
\end{figure}

\begin{sidewaystable}[pt!]
	\centering
	\caption{Differences in SD p.a. of GMV-50SP across different estimators.\label{table:sign_50sp}}
	\resizebox{\linewidth}{!}{%
		\begin{threeparttable}
			\begin{tabular}{l *{11}{d{2.6}}}
				\toprule
				& \mc{Sample} &\mc{LW\textsubscript{1}} & \mc{LW\textsubscript{1}-CV2} & \mc{LW\textsubscript{CC}} & \mc{LW\textsubscript{CC}-CV2} & \mc{LW\textsubscript{NL}} & \mc{LW\textsubscript{NL}-CV2} & \mc{POET} & \mc{POET-CV2} & \mc{GLASSO} & \mc{GLASSO-CV2} \\
				\midrule
				Sample &  & \cellcolor[rgb]{0.186,0.862,0.169}-0.00145^{***} & \cellcolor[rgb]{0.119,0.84,0.18}-0.00163^* & \cellcolor[rgb]{0,0.8,0.2}-0.00210^{***} & \cellcolor[rgb]{0,0.8,0.2}-0.00207^* & \cellcolor[rgb]{0.033,0.811,0.194}-0.00186^{***} & \cellcolor[rgb]{0.033,0.811,0.194}-0.00186^{***} & \cellcolor[rgb]{0,0.8,0.2}-0.00262^{***} & \cellcolor[rgb]{0,0.8,0.2}-0.00257^{***} & \cellcolor[rgb]{0,0.8,0.2}-0.00229^{**} & \cellcolor[rgb]{0,0.8,0.2}-0.00441^{***} \\ 
				LW\textsubscript{1} & \cellcolor[rgb]{1,0.48,0}0.00145 &  & \cellcolor[rgb]{0.829,1,0.043}-0.00018 & \cellcolor[rgb]{0.484,0.961,0.119}-0.00065 & \cellcolor[rgb]{0.497,0.966,0.117}-0.00062 & \cellcolor[rgb]{0.576,0.992,0.104}-0.00041^* & \cellcolor[rgb]{0.576,0.992,0.104}-0.00041 & \cellcolor[rgb]{0.291,0.897,0.152}-0.00117^{***} & \cellcolor[rgb]{0.308,0.903,0.149}-0.00112^{***} & \cellcolor[rgb]{0.412,0.937,0.131}-0.00085 & \cellcolor[rgb]{0,0.8,0.2}-0.00296^{***} \\ 
				LW\textsubscript{1}-CV2 & \cellcolor[rgb]{1,0.427,0}0.00163 & \cellcolor[rgb]{0.903,0.897,0.194}0.00018 &  & \cellcolor[rgb]{0.552,0.984,0.108}-0.00047 & \cellcolor[rgb]{0.564,0.988,0.106}-0.00044 & \cellcolor[rgb]{0.763,1,0.059}-0.00023 & \cellcolor[rgb]{0.763,1,0.059}-0.00023 & \cellcolor[rgb]{0.359,0.92,0.14}-0.00099 & \cellcolor[rgb]{0.376,0.925,0.137}-0.00094 & \cellcolor[rgb]{0.48,0.96,0.12}-0.00067 & \cellcolor[rgb]{0,0.8,0.2}-0.00278^{***} \\ 
				LW\textsubscript{CC} & \cellcolor[rgb]{1,0.289,0}0.00210 & \cellcolor[rgb]{1,0.712,0}0.00065 & \cellcolor[rgb]{1,0.764,0}0.00047 &  & \cellcolor[rgb]{0.818,0.982,0.363}0.00003 & \cellcolor[rgb]{0.94,0.86,0.12}0.00024 & \cellcolor[rgb]{0.94,0.86,0.121}0.00024 & \cellcolor[rgb]{0.536,0.979,0.111}-0.00052 & \cellcolor[rgb]{0.553,0.984,0.108}-0.00047 & \cellcolor[rgb]{0.812,1,0.047}-0.00019 & \cellcolor[rgb]{0,0.8,0.2}-0.00231^{***} \\ 
				LW\textsubscript{CC}-CV2 & \cellcolor[rgb]{1,0.299,0}0.00207 & \cellcolor[rgb]{1,0.721,0}0.00062 & \cellcolor[rgb]{1,0.774,0}0.00044 & \cellcolor[rgb]{0.913,1,0.174}-0.00003 &  & \cellcolor[rgb]{0.921,0.879,0.157}0.00021 & \cellcolor[rgb]{0.921,0.879,0.158}0.00021 & \cellcolor[rgb]{0.524,0.975,0.113}-0.00055 & \cellcolor[rgb]{0.541,0.98,0.11}-0.00050 & \cellcolor[rgb]{0.767,1,0.058}-0.00023 & \cellcolor[rgb]{0,0.8,0.2}-0.00234^{***} \\ 
				LW\textsubscript{NL} & \cellcolor[rgb]{1,0.361,0}0.00186 & \cellcolor[rgb]{1,0.783,0}0.00041 & \cellcolor[rgb]{0.93,0.87,0.14}0.00023 & \cellcolor[rgb]{0.739,1,0.065}-0.00024 & \cellcolor[rgb]{0.784,1,0.054}-0.00021 &  & \cellcolor[rgb]{0.8,1,0.399}-0.00000 & \cellcolor[rgb]{0.444,0.948,0.126}-0.00076^* & \cellcolor[rgb]{0.462,0.954,0.123}-0.00071^* & \cellcolor[rgb]{0.565,0.988,0.106}-0.00044 & \cellcolor[rgb]{0,0.8,0.2}-0.00255^{***} \\ 
				LW\textsubscript{NL}-CV2 & \cellcolor[rgb]{1,0.361,0}0.00186 & \cellcolor[rgb]{1,0.783,0}0.00041 & \cellcolor[rgb]{0.93,0.87,0.14}0.00023 & \cellcolor[rgb]{0.739,1,0.065}-0.00024 & \cellcolor[rgb]{0.784,1,0.054}-0.00021 & \cellcolor[rgb]{0.8,1,0.4}0.00000 &  & \cellcolor[rgb]{0.444,0.948,0.126}-0.00076^* & \cellcolor[rgb]{0.462,0.954,0.123}-0.00071^* & \cellcolor[rgb]{0.565,0.988,0.106}-0.00044 & \cellcolor[rgb]{0,0.8,0.2}-0.00255^{***} \\ 
				POET & \cellcolor[rgb]{1,0.27,0}0.00262 & \cellcolor[rgb]{1,0.561,0}0.00117 & \cellcolor[rgb]{1,0.614,0}0.00099 & \cellcolor[rgb]{1,0.752,0}0.00052 & \cellcolor[rgb]{1,0.742,0}0.00055 & \cellcolor[rgb]{1,0.68,0}0.00076 & \cellcolor[rgb]{1,0.68,0}0.00076 &  & \cellcolor[rgb]{0.827,0.973,0.347}0.00005 & \cellcolor[rgb]{0.985,0.815,0.031}0.00032 & \cellcolor[rgb]{0.059,0.82,0.19}-0.00179^{**} \\ 
				POET-CV2 & \cellcolor[rgb]{1,0.27,0}0.00257 & \cellcolor[rgb]{1,0.575,0}0.00112 & \cellcolor[rgb]{1,0.627,0}0.00094 & \cellcolor[rgb]{1,0.765,0}0.00047 & \cellcolor[rgb]{1,0.756,0}0.00050 & \cellcolor[rgb]{1,0.694,0}0.00071 & \cellcolor[rgb]{1,0.694,0}0.00071 & \cellcolor[rgb]{0.963,1,0.074}-0.00005 &  & \cellcolor[rgb]{0.958,0.842,0.084}0.00028 & \cellcolor[rgb]{0.041,0.814,0.193}-0.00184^{**} \\ 
				GLASSO & \cellcolor[rgb]{1,0.27,0}0.00229 & \cellcolor[rgb]{1,0.655,0}0.00085 & \cellcolor[rgb]{1,0.708,0}0.00067 & \cellcolor[rgb]{0.91,0.89,0.18}0.00019 & \cellcolor[rgb]{0.929,0.871,0.143}0.00023 & \cellcolor[rgb]{1,0.774,0}0.00044 & \cellcolor[rgb]{1,0.775,0}0.00044 & \cellcolor[rgb]{0.63,1,0.093}-0.00032 & \cellcolor[rgb]{0.695,1,0.076}-0.00028 &  & \cellcolor[rgb]{0,0.8,0.2}-0.00211^{**} \\ 
				GLASSO-CV2 & \cellcolor[rgb]{1,0.27,0}0.00441 & \cellcolor[rgb]{1,0.27,0}0.00296 & \cellcolor[rgb]{1,0.27,0}0.00278 & \cellcolor[rgb]{1,0.27,0}0.00231 & \cellcolor[rgb]{1,0.27,0}0.00234 & \cellcolor[rgb]{1,0.27,0}0.00255 & \cellcolor[rgb]{1,0.27,0}0.00255 & \cellcolor[rgb]{1,0.38,0}0.00179 & \cellcolor[rgb]{1,0.367,0}0.00184 & \cellcolor[rgb]{1,0.286,0}0.00211 &  \\ \midrule
				better than \% of models & 0.0 & 0.1 & 0.2 & 0.6 & 0.5 & 0.3 & 0.4 & 0.9 & 0.8 & 0.7 & 1.0 \\ 
				\bottomrule
			\end{tabular}
			\begin{tablenotes}
				\item This table shows the differences in the annualized out-of-sample SD of the GMV-50SP portfolios across the main covariance estimation methods and their CV2-based counterparts. The table is constructed in a symmetrical way with an applied color scheme from red (higher SD than the other model) to green (lower SD than the other model). In addition, on the elements above the diagonal, the significant pairwise outperformance in terms of variance is denoted by asterisks: *** denotes significance at the 0.001 level; ** denotes significance at the 0.01 level; and * denotes significance at the 0.05 level. Finally, for each model, we report the percentage of the other models that exhibit higher variance as a qualitative measure.  
			\end{tablenotes}
		\end{threeparttable}
	}
	\vspace{1cm} 
	\caption{Differences in SD p.a. of GMV-100SP across different estimators.\label{table:sign_100sp}}
	\resizebox{\linewidth}{!}{%
		\begin{threeparttable}
			\begin{tabular}{l *{11}{d{2.6}}}
				\toprule
				& \mc{Sample} &\mc{LW\textsubscript{1}} & \mc{LW\textsubscript{1}-CV2} & \mc{LW\textsubscript{CC}} & \mc{LW\textsubscript{CC}-CV2} & \mc{LW\textsubscript{NL}} & \mc{LW\textsubscript{NL}-CV2} & \mc{POET} & \mc{POET-CV2} & \mc{GLASSO} & \mc{GLASSO-CV2} \\
				\midrule
				Sample &  & \cellcolor[rgb]{0.156,0.852,0.174}-0.00310^{***} & \cellcolor[rgb]{0,0.8,0.2}-0.00502^{***} & \cellcolor[rgb]{0.134,0.845,0.178}-0.00323^{***} & \cellcolor[rgb]{0,0.8,0.2}-0.00430^{***} & \cellcolor[rgb]{0,0.8,0.2}-0.00450^{***} & \cellcolor[rgb]{0,0.8,0.2}-0.00451^{***} & \cellcolor[rgb]{0,0.8,0.2}-0.00433^{***} & \cellcolor[rgb]{0,0.8,0.2}-0.00484^{***} & \cellcolor[rgb]{0,0.8,0.2}-0.00576^{***} & \cellcolor[rgb]{0,0.8,0.2}-0.00748^{***} \\ 
				LW\textsubscript{1} & \cellcolor[rgb]{1,0.27,0}0.00310 &  & \cellcolor[rgb]{0.364,0.921,0.139}-0.00191^{**} & \cellcolor[rgb]{0.757,1,0.335}-0.00012 & \cellcolor[rgb]{0.489,0.963,0.118}-0.00120 & \cellcolor[rgb]{0.454,0.951,0.124}-0.00140^{***} & \cellcolor[rgb]{0.453,0.951,0.124}-0.00140^{***} & \cellcolor[rgb]{0.485,0.962,0.119}-0.00122^{***} & \cellcolor[rgb]{0.395,0.932,0.134}-0.00173^{***} & \cellcolor[rgb]{0.234,0.878,0.161}-0.00266^{***} & \cellcolor[rgb]{0,0.8,0.2}-0.00438^{***} \\ 
				LW\textsubscript{1}-CV2 & \cellcolor[rgb]{1,0.27,0}0.00502 & \cellcolor[rgb]{1,0.519,0}0.00191 &  & \cellcolor[rgb]{1,0.545,0}0.00179^* & \cellcolor[rgb]{1,0.78,0}0.00072 & \cellcolor[rgb]{1,0.85,0}0.00051 & \cellcolor[rgb]{1,0.852,0}0.00051 & \cellcolor[rgb]{1,0.786,0}0.00069 & \cellcolor[rgb]{1,0.999,0}0.00018 & \cellcolor[rgb]{0.568,0.989,0.105}-0.00074 & \cellcolor[rgb]{0.267,0.889,0.155}-0.00247^{***} \\ 
				LW\textsubscript{CC} & \cellcolor[rgb]{1,0.27,0}0.00323 & \cellcolor[rgb]{0.938,1,0.124}0.00012 & \cellcolor[rgb]{0.385,0.928,0.136}-0.00179 &  & \cellcolor[rgb]{0.511,0.97,0.115}-0.00107^* & \cellcolor[rgb]{0.475,0.958,0.121}-0.00128^{**} & \cellcolor[rgb]{0.474,0.958,0.121}-0.00128^{**} & \cellcolor[rgb]{0.506,0.969,0.116}-0.00110^{**} & \cellcolor[rgb]{0.417,0.939,0.131}-0.00161^{***} & \cellcolor[rgb]{0.255,0.885,0.157}-0.00253^{***} & \cellcolor[rgb]{0,0.8,0.2}-0.00426^{***} \\ 
				LW\textsubscript{CC}-CV2 & \cellcolor[rgb]{1,0.27,0}0.00430 & \cellcolor[rgb]{1,0.675,0}0.00120 & \cellcolor[rgb]{0.573,0.991,0.104}-0.00072 & \cellcolor[rgb]{1,0.702,0}0.00107 &  & \cellcolor[rgb]{0.728,1,0.292}-0.00020 & \cellcolor[rgb]{0.727,1,0.29}-0.00021 & \cellcolor[rgb]{0.791,1,0.387}-0.00002 & \cellcolor[rgb]{0.609,1,0.113}-0.00054 & \cellcolor[rgb]{0.443,0.948,0.126}-0.00146^* & \cellcolor[rgb]{0.142,0.847,0.176}-0.00318^{***} \\ 
				LW\textsubscript{NL} & \cellcolor[rgb]{1,0.27,0}0.00450 & \cellcolor[rgb]{1,0.631,0}0.00140 & \cellcolor[rgb]{0.617,1,0.126}-0.00051 & \cellcolor[rgb]{1,0.658,0}0.00128 & \cellcolor[rgb]{1,0.989,0}0.00020 &  & \cellcolor[rgb]{0.799,1,0.398}-0.00000 & \cellcolor[rgb]{1,0.999,0}0.00018 & \cellcolor[rgb]{0.681,1,0.221}-0.00033 & \cellcolor[rgb]{0.478,0.959,0.12}-0.00126^* & \cellcolor[rgb]{0.178,0.859,0.17}-0.00298^{***} \\ 
				LW\textsubscript{NL}-CV2 & \cellcolor[rgb]{1,0.27,0}0.00451 & \cellcolor[rgb]{1,0.63,0}0.00140 & \cellcolor[rgb]{0.619,1,0.128}-0.00051 & \cellcolor[rgb]{1,0.657,0}0.00128 & \cellcolor[rgb]{1,0.987,0}0.00021 & \cellcolor[rgb]{0.804,1,0.392}0.00000 &  & \cellcolor[rgb]{1,0.998,0}0.00018 & \cellcolor[rgb]{0.682,1,0.223}-0.00033 & \cellcolor[rgb]{0.479,0.96,0.12}-0.00125^* & \cellcolor[rgb]{0.178,0.859,0.17}-0.00298^{***} \\ 
				POET & \cellcolor[rgb]{1,0.27,0}0.00433 & \cellcolor[rgb]{1,0.67,0}0.00122 & \cellcolor[rgb]{0.577,0.992,0.104}-0.00069 & \cellcolor[rgb]{1,0.697,0}0.00110 & \cellcolor[rgb]{0.828,1,0.345}0.00002 & \cellcolor[rgb]{0.737,1,0.305}-0.00018 & \cellcolor[rgb]{0.735,1,0.303}-0.00018 &  & \cellcolor[rgb]{0.617,1,0.126}-0.00051^* & \cellcolor[rgb]{0.447,0.949,0.125}-0.00144^* & \cellcolor[rgb]{0.147,0.849,0.176}-0.00316^{***} \\ 
				POET-CV2 & \cellcolor[rgb]{1,0.27,0}0.00484 & \cellcolor[rgb]{1,0.558,0}0.00173 & \cellcolor[rgb]{0.737,1,0.305}-0.00018 & \cellcolor[rgb]{1,0.584,0}0.00161 & \cellcolor[rgb]{1,0.84,0}0.00054 & \cellcolor[rgb]{1,0.93,0}0.00033 & \cellcolor[rgb]{1,0.931,0}0.00033 & \cellcolor[rgb]{1,0.85,0}0.00051 &  & \cellcolor[rgb]{0.537,0.979,0.111}-0.00092 & \cellcolor[rgb]{0.236,0.879,0.161}-0.00264^{***} \\ 
				GLASSO & \cellcolor[rgb]{1,0.27,0}0.00576 & \cellcolor[rgb]{1,0.356,0}0.00266 & \cellcolor[rgb]{1,0.774,0}0.00074 & \cellcolor[rgb]{1,0.382,0}0.00253 & \cellcolor[rgb]{1,0.617,0}0.00146 & \cellcolor[rgb]{1,0.662,0}0.00126 & \cellcolor[rgb]{1,0.663,0}0.00125 & \cellcolor[rgb]{1,0.623,0}0.00144 & \cellcolor[rgb]{1,0.735,0}0.00092 &  & \cellcolor[rgb]{0.397,0.932,0.134}-0.00172^{***} \\ 
				GLASSO-CV2 & \cellcolor[rgb]{1,0.27,0}0.00748 & \cellcolor[rgb]{1,0.27,0}0.00438 & \cellcolor[rgb]{1,0.397,0}0.00247 & \cellcolor[rgb]{1,0.27,0}0.00426 & \cellcolor[rgb]{1,0.27,0}0.00318 & \cellcolor[rgb]{1,0.285,0}0.00298 & \cellcolor[rgb]{1,0.286,0}0.00298 & \cellcolor[rgb]{1,0.27,0}0.00316 & \cellcolor[rgb]{1,0.358,0}0.00264 & \cellcolor[rgb]{1,0.56,0}0.00172 &  \\ 	\toprule
				better than \% of models & 0.0 & 0.1 & 0.8 & 0.2 & 0.3 & 0.5 & 0.6 & 0.4 & 0.7 & 0.9 & 1.0 \\ 
				\bottomrule
			\end{tabular}
			\begin{tablenotes}
				\item This table shows the differences in the annualized out-of-sample SD of the GMV-100SP portfolios across the main covariance estimation methods and their CV2-based counterparts. The table is constructed in a symmetrical way with an applied color scheme from red (higher SD than the other model) to green (lower SD than the other model). In addition, on the elements above the diagonal, the significant pairwise outperformance in terms of variance is denoted by asterisks: *** denotes significance at the 0.001 level; ** denotes significance at the 0.01 level; and * denotes significance at the 0.05 level. Finally, for each model, we report the percentage of the other models that exhibit higher variance as a qualitative measure.   
			\end{tablenotes}
		\end{threeparttable}
	}
\end{sidewaystable}

\begin{sidewaystable}[pt!]
	\centering
	\caption{Differences in SD p.a. of GMV-200SP across different estimators.\label{table:sign_200sp}}
	\resizebox{\linewidth}{!}{%
		\begin{threeparttable}
			\begin{tabular}{l *{11}{d{2.6}}}
				\toprule
				& \mc{Sample} &\mc{LW\textsubscript{1}} & \mc{LW\textsubscript{1}-CV2} & \mc{LW\textsubscript{CC}} & \mc{LW\textsubscript{CC}-CV2} & \mc{LW\textsubscript{NL}} & \mc{LW\textsubscript{NL}-CV2} & \mc{POET} & \mc{POET-CV2} & \mc{GLASSO} & \mc{GLASSO-CV2} \\
				\midrule
				Sample &  & \cellcolor[rgb]{0.049,0.816,0.192}-0.00887^{***} & \cellcolor[rgb]{0,0.8,0.2}-0.01275^{***} & \cellcolor[rgb]{0.05,0.817,0.192}-0.00885^{***} & \cellcolor[rgb]{0,0.8,0.2}-0.01049^{***} & \cellcolor[rgb]{0,0.8,0.2}-0.01277^{***} & \cellcolor[rgb]{0,0.8,0.2}-0.01276^{***} & \cellcolor[rgb]{0,0.8,0.2}-0.01103^{***} & \cellcolor[rgb]{0,0.8,0.2}-0.01080^{***} & \cellcolor[rgb]{0,0.8,0.2}-0.01264^{***} & \cellcolor[rgb]{0,0.8,0.2}-0.01342^{***} \\ 
				LW\textsubscript{1} & \cellcolor[rgb]{1,0.27,0}0.00887 &  & \cellcolor[rgb]{0.388,0.929,0.135}-0.00388^{***} & \cellcolor[rgb]{0.803,1,0.394}0.00002 & \cellcolor[rgb]{0.542,0.981,0.11}-0.00162 & \cellcolor[rgb]{0.387,0.929,0.136}-0.00390^{***} & \cellcolor[rgb]{0.388,0.929,0.135}-0.00389^{***} & \cellcolor[rgb]{0.505,0.968,0.116}-0.00216^{***} & \cellcolor[rgb]{0.521,0.974,0.113}-0.00193^{***} & \cellcolor[rgb]{0.396,0.932,0.134}-0.00377^{**} & \cellcolor[rgb]{0.343,0.914,0.143}-0.00455^{***} \\ 
				LW\textsubscript{1}-CV2 & \cellcolor[rgb]{1,0.27,0}0.01275 & \cellcolor[rgb]{1,0.27,0}0.00388 &  & \cellcolor[rgb]{1,0.27,0}0.00390^{***} & \cellcolor[rgb]{1,0.812,0}0.00226^* & \cellcolor[rgb]{0.794,1,0.391}-0.00002 & \cellcolor[rgb]{0.798,1,0.397}-0.00001 & \cellcolor[rgb]{1,0.894,0}0.00172 & \cellcolor[rgb]{1,0.86,0}0.00195^* & \cellcolor[rgb]{0.822,1,0.356}0.00011 & \cellcolor[rgb]{0.624,1,0.136}-0.00067 \\ 
				LW\textsubscript{CC} & \cellcolor[rgb]{1,0.27,0}0.00885 & \cellcolor[rgb]{0.796,1,0.394}-0.00002 & \cellcolor[rgb]{0.387,0.929,0.135}-0.00390 &  & \cellcolor[rgb]{0.541,0.98,0.11}-0.00164^{**} & \cellcolor[rgb]{0.386,0.929,0.136}-0.00392^{***} & \cellcolor[rgb]{0.387,0.929,0.136}-0.00390^{***} & \cellcolor[rgb]{0.504,0.968,0.116}-0.00218^{***} & \cellcolor[rgb]{0.519,0.973,0.113}-0.00195^{**} & \cellcolor[rgb]{0.395,0.932,0.134}-0.00379^{**} & \cellcolor[rgb]{0.341,0.914,0.143}-0.00457^{***} \\ 
				LW\textsubscript{CC}-CV2 & \cellcolor[rgb]{1,0.27,0}0.01049 & \cellcolor[rgb]{1,0.909,0}0.00162 & \cellcolor[rgb]{0.498,0.966,0.117}-0.00226 & \cellcolor[rgb]{1,0.906,0}0.00164 &  & \cellcolor[rgb]{0.497,0.966,0.117}-0.00228^{**} & \cellcolor[rgb]{0.498,0.966,0.117}-0.00227^{**} & \cellcolor[rgb]{0.658,1,0.187}-0.00054 & \cellcolor[rgb]{0.718,1,0.277}-0.00031 & \cellcolor[rgb]{0.506,0.969,0.116}-0.00215^* & \cellcolor[rgb]{0.453,0.951,0.125}-0.00293^{**} \\ 
				LW\textsubscript{NL} & \cellcolor[rgb]{1,0.27,0}0.01277 & \cellcolor[rgb]{1,0.27,0}0.00390 & \cellcolor[rgb]{0.805,1,0.391}0.00002 & \cellcolor[rgb]{1,0.27,0}0.00392 & \cellcolor[rgb]{1,0.809,0}0.00228 &  & \cellcolor[rgb]{0.803,1,0.394}0.00002 & \cellcolor[rgb]{1,0.89,0}0.00174^{**} & \cellcolor[rgb]{1,0.856,0}0.00197^{***} & \cellcolor[rgb]{0.827,1,0.347}0.00013 & \cellcolor[rgb]{0.63,1,0.145}-0.00065 \\ 
				LW\textsubscript{NL}-CV2 & \cellcolor[rgb]{1,0.27,0}0.01276 & \cellcolor[rgb]{1,0.27,0}0.00389 & \cellcolor[rgb]{0.801,1,0.397}0.00001 & \cellcolor[rgb]{1,0.27,0}0.00390 & \cellcolor[rgb]{1,0.811,0}0.00227 & \cellcolor[rgb]{0.796,1,0.394}-0.00002 &  & \cellcolor[rgb]{1,0.893,0}0.00173^{**} & \cellcolor[rgb]{1,0.859,0}0.00195^{***} & \cellcolor[rgb]{0.823,1,0.353}0.00012 & \cellcolor[rgb]{0.626,1,0.139}-0.00067 \\ 
				POET & \cellcolor[rgb]{1,0.27,0}0.01103 & \cellcolor[rgb]{1,0.827,0}0.00216 & \cellcolor[rgb]{0.535,0.978,0.111}-0.00172 & \cellcolor[rgb]{1,0.825,0}0.00218 & \cellcolor[rgb]{0.908,1,0.185}0.00054 & \cellcolor[rgb]{0.534,0.978,0.111}-0.00174 & \cellcolor[rgb]{0.535,0.978,0.111}-0.00173 &  & \cellcolor[rgb]{0.845,1,0.31}0.00023 & \cellcolor[rgb]{0.543,0.981,0.11}-0.00161 & \cellcolor[rgb]{0.489,0.963,0.118}-0.00239^* \\ 
				POET-CV2 & \cellcolor[rgb]{1,0.27,0}0.01080 & \cellcolor[rgb]{1,0.862,0}0.00193 & \cellcolor[rgb]{0.52,0.973,0.113}-0.00195 & \cellcolor[rgb]{1,0.859,0}0.00195 & \cellcolor[rgb]{0.862,1,0.275}0.00031 & \cellcolor[rgb]{0.518,0.973,0.114}-0.00197 & \cellcolor[rgb]{0.519,0.973,0.113}-0.00195 & \cellcolor[rgb]{0.74,1,0.311}-0.00023 &  & \cellcolor[rgb]{0.527,0.976,0.112}-0.00184 & \cellcolor[rgb]{0.474,0.958,0.121}-0.00262^* \\ 
				GLASSO & \cellcolor[rgb]{1,0.27,0}0.01264 & \cellcolor[rgb]{1,0.27,0}0.00377 & \cellcolor[rgb]{0.771,1,0.356}-0.00011 & \cellcolor[rgb]{1,0.27,0}0.00379 & \cellcolor[rgb]{1,0.829,0}0.00215 & \cellcolor[rgb]{0.765,1,0.347}-0.00013 & \cellcolor[rgb]{0.769,1,0.354}-0.00012 & \cellcolor[rgb]{1,0.91,0}0.00161 & \cellcolor[rgb]{1,0.876,0}0.00184 &  & \cellcolor[rgb]{0.599,1,0.1}-0.00078^* \\ 
				GLASSO-CV2 & \cellcolor[rgb]{1,0.27,0}0.01342 & \cellcolor[rgb]{1,0.27,0}0.00455 & \cellcolor[rgb]{0.934,1,0.133}0.00067 & \cellcolor[rgb]{1,0.27,0}0.00457 & \cellcolor[rgb]{1,0.296,0}0.00293 & \cellcolor[rgb]{0.929,1,0.142}0.00065 & \cellcolor[rgb]{0.932,1,0.136}0.00067 & \cellcolor[rgb]{1,0.76,0}0.00239 & \cellcolor[rgb]{1,0.565,0}0.00262 & \cellcolor[rgb]{0.956,1,0.089}0.00078 &  \\ \midrule
				better than \% of models & 0.0 & 0.2 & 0.7 & 0.1 & 0.3 & 0.9 & 0.8 & 0.5 & 0.4 & 0.6 & 1.0 \\ 
				\bottomrule
			\end{tabular}
			\begin{tablenotes}
				\item This table shows the differences in the annualized out-of-sample SD of the GMV-200SP portfolios across the main covariance estimation methods and their CV2-based counterparts. The table is constructed in a symmetrical way with an applied color scheme from red (higher SD than the other model) to green (lower SD than the other model). In addition, on the elements above the diagonal, the significant pairwise outperformance in terms of variance is denoted by asterisks: *** denotes significance at the 0.001 level; ** denotes significance at the 0.01 level; and * denotes significance at the 0.05 level. Finally, for each model, we report the percentage of the other models that exhibit higher variance as a qualitative measure.
			\end{tablenotes}
		\end{threeparttable}
	}
\end{sidewaystable}

\clearpage
\section{GMV without short sales}\label{app:gmv_noshort}

\begin{figure}[h]
	\begin{center}
		\includegraphics[width=\textwidth]{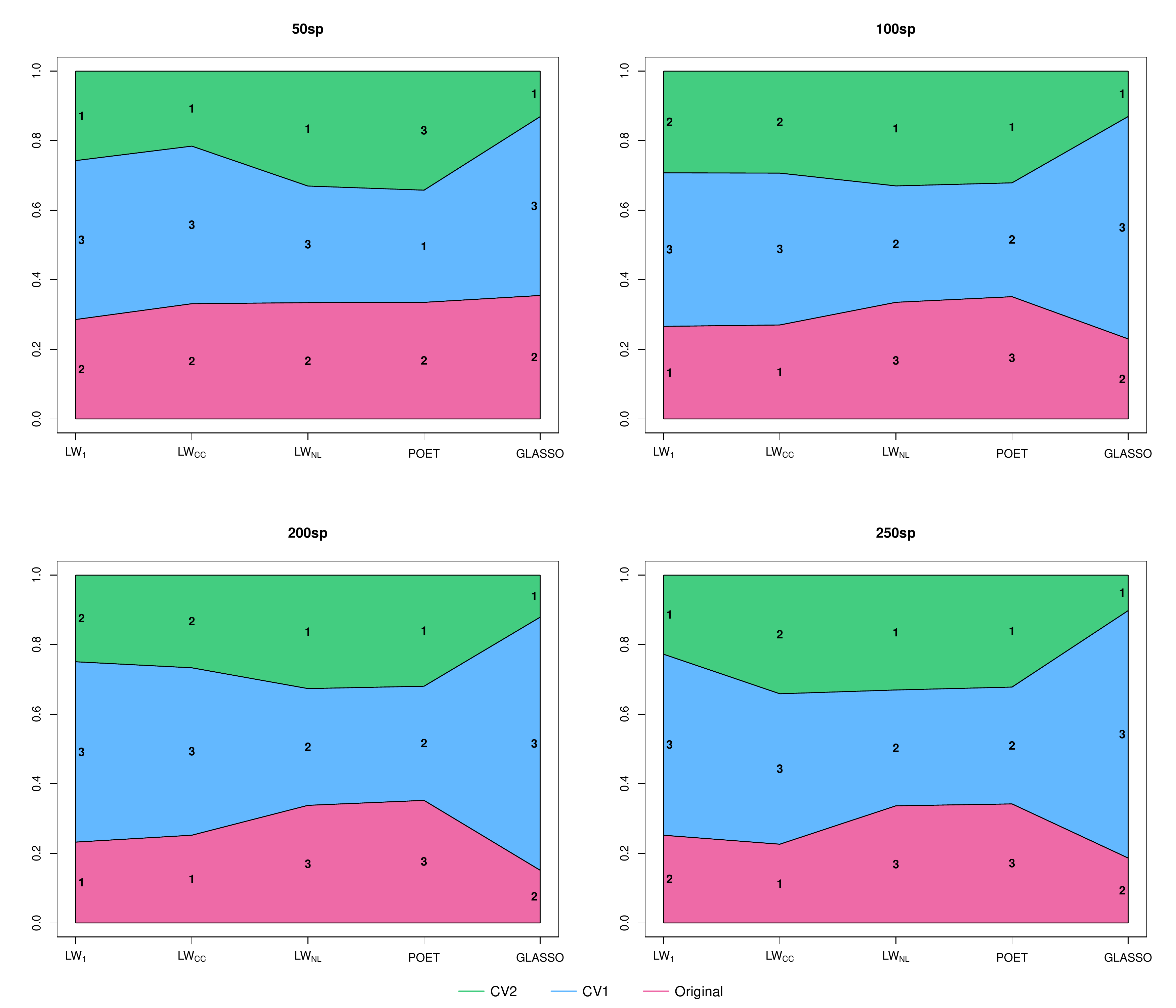}
		\caption{Relative differences in annualized SD of GMV-NOSHORT across all the covariance estimators.}\label{fig:sdcomparison_noshort}
	\end{center}
\end{figure}

\begin{sidewaystable}[pt!]
	\centering
	\caption{Differences in SD p.a. of GMV-NOSHORT-50SP across different estimators.\label{table:sign_50sp_noshort}}
	\resizebox{\linewidth}{!}{%
		\begin{threeparttable}
			\begin{tabular}{l *{11}{d{2.6}}}
				\toprule
				& \mc{Sample} &\mc{LW\textsubscript{1}} & \mc{LW\textsubscript{1}-CV2} & \mc{LW\textsubscript{CC}} & \mc{LW\textsubscript{CC}-CV2} & \mc{LW\textsubscript{NL}} & \mc{LW\textsubscript{NL}-CV2} & \mc{POET} & \mc{POET-CV2} & \mc{GLASSO} & \mc{GLASSO-CV2} \\
				\midrule
				Sample &  & \cellcolor[rgb]{0.842,0.958,0.317}0.00004 & \cellcolor[rgb]{0.783,1,0.054}-0.00017 & \cellcolor[rgb]{0.46,0.953,0.123}-0.00040 & \cellcolor[rgb]{0,0.8,0.2}-0.00097^{**} & \cellcolor[rgb]{0.819,0.981,0.362}0.00002 & \cellcolor[rgb]{0.819,1,0.363}-0.00001 & \cellcolor[rgb]{0.404,0.935,0.133}-0.00045^{**} & \cellcolor[rgb]{0.439,0.946,0.127}-0.00042^{**} & \cellcolor[rgb]{1,0.383,0}0.00066 & \cellcolor[rgb]{0,0.8,0.2}-0.00105^{**} \\ 
				LW\textsubscript{1} & \cellcolor[rgb]{0.923,1,0.154}-0.00004 &  & \cellcolor[rgb]{0.704,1,0.074}-0.00021 & \cellcolor[rgb]{0.421,0.94,0.13}-0.00044 & \cellcolor[rgb]{0,0.8,0.2}-0.00101^* & \cellcolor[rgb]{0.868,1,0.265}-0.00002 & \cellcolor[rgb]{0.942,1,0.116}-0.00004 & \cellcolor[rgb]{0.365,0.922,0.139}-0.00049^* & \cellcolor[rgb]{0.4,0.933,0.133}-0.00046^* & \cellcolor[rgb]{1,0.417,0}0.00062 & \cellcolor[rgb]{0,0.8,0.2}-0.00109^{***} \\ 
				LW\textsubscript{1}-CV2 & \cellcolor[rgb]{0.982,0.818,0.037}0.00017 & \cellcolor[rgb]{1,0.781,0}0.00021 &  & \cellcolor[rgb]{0.661,1,0.085}-0.00023 & \cellcolor[rgb]{0.051,0.817,0.191}-0.00080^* & \cellcolor[rgb]{1,0.8,0}0.00019 & \cellcolor[rgb]{0.975,0.825,0.05}0.00016 & \cellcolor[rgb]{0.574,0.991,0.104}-0.00028 & \cellcolor[rgb]{0.618,1,0.096}-0.00025 & \cellcolor[rgb]{1,0.27,0}0.00083 & \cellcolor[rgb]{0,0.8,0.2}-0.00088^{**} \\ 
				LW\textsubscript{CC} & \cellcolor[rgb]{1,0.614,0}0.00040 & \cellcolor[rgb]{1,0.58,0}0.00044 & \cellcolor[rgb]{1,0.762,0}0.00023 &  & \cellcolor[rgb]{0.282,0.894,0.153}-0.00057^{**} & \cellcolor[rgb]{1,0.599,0}0.00042 & \cellcolor[rgb]{1,0.62,0}0.00039 & \cellcolor[rgb]{0.976,1,0.048}-0.00006 & \cellcolor[rgb]{0.867,1,0.266}-0.00002 & \cellcolor[rgb]{1,0.27,0}0.00106 & \cellcolor[rgb]{0.207,0.869,0.166}-0.00065^* \\ 
				LW\textsubscript{CC}-CV2 & \cellcolor[rgb]{1,0.27,0}0.00097 & \cellcolor[rgb]{1,0.27,0}0.00101 & \cellcolor[rgb]{1,0.27,0}0.00080 & \cellcolor[rgb]{1,0.459,0}0.00057 &  & \cellcolor[rgb]{1,0.27,0}0.00099^* & \cellcolor[rgb]{1,0.27,0}0.00097^* & \cellcolor[rgb]{1,0.508,0}0.00052 & \cellcolor[rgb]{1,0.478,0}0.00055 & \cellcolor[rgb]{1,0.27,0}0.00163^{**} & \cellcolor[rgb]{0.976,1,0.006}-0.00007 \\ 
				LW\textsubscript{NL} & \cellcolor[rgb]{0.856,1,0.289}-0.00002 & \cellcolor[rgb]{0.823,0.977,0.354}0.00002 & \cellcolor[rgb]{0.748,1,0.063}-0.00019 & \cellcolor[rgb]{0.442,0.947,0.126}-0.00042 & \cellcolor[rgb]{0,0.8,0.2}-0.00099 &  & \cellcolor[rgb]{0.874,1,0.251}-0.00002 & \cellcolor[rgb]{0.387,0.929,0.136}-0.00047^* & \cellcolor[rgb]{0.421,0.94,0.13}-0.00044^* & \cellcolor[rgb]{1,0.398,0}0.00064 & \cellcolor[rgb]{0,0.8,0.2}-0.00106^{***} \\ 
				LW\textsubscript{NL}-CV2 & \cellcolor[rgb]{0.806,0.994,0.387}0.00001 & \cellcolor[rgb]{0.848,0.952,0.304}0.00004 & \cellcolor[rgb]{0.796,1,0.051}-0.00016 & \cellcolor[rgb]{0.466,0.955,0.122}-0.00039 & \cellcolor[rgb]{0,0.8,0.2}-0.00097 & \cellcolor[rgb]{0.825,0.975,0.35}0.00002 &  & \cellcolor[rgb]{0.41,0.937,0.132}-0.00045^* & \cellcolor[rgb]{0.445,0.948,0.126}-0.00041 & \cellcolor[rgb]{1,0.378,0}0.00067 & \cellcolor[rgb]{0,0.8,0.2}-0.00104^{**} \\ 
				POET & \cellcolor[rgb]{1,0.566,0}0.00045 & \cellcolor[rgb]{1,0.532,0}0.00049 & \cellcolor[rgb]{1,0.714,0}0.00028 & \cellcolor[rgb]{0.86,0.94,0.281}0.00006 & \cellcolor[rgb]{0.337,0.912,0.144}-0.00052 & \cellcolor[rgb]{1,0.551,0}0.00047 & \cellcolor[rgb]{1,0.571,0}0.00045 &  & \cellcolor[rgb]{0.837,0.963,0.326}0.00003 & \cellcolor[rgb]{1,0.27,0}0.00111 & \cellcolor[rgb]{0.263,0.888,0.156}-0.00059 \\ 
				POET-CV2 & \cellcolor[rgb]{1,0.596,0}0.00042 & \cellcolor[rgb]{1,0.562,0}0.00046 & \cellcolor[rgb]{1,0.744,0}0.00025 & \cellcolor[rgb]{0.823,0.977,0.355}0.00002 & \cellcolor[rgb]{0.303,0.901,0.15}-0.00055 & \cellcolor[rgb]{1,0.581,0}0.00044 & \cellcolor[rgb]{1,0.601,0}0.00041 & \cellcolor[rgb]{0.909,1,0.182}-0.00003 &  & \cellcolor[rgb]{1,0.27,0}0.00108 & \cellcolor[rgb]{0.228,0.876,0.162}-0.00063^* \\ 
				GLASSO & \cellcolor[rgb]{0.194,0.865,0.168}-0.00066 & \cellcolor[rgb]{0.233,0.878,0.161}-0.00062 & \cellcolor[rgb]{0.024,0.808,0.196}-0.00083 & \cellcolor[rgb]{0,0.8,0.2}-0.00106 & \cellcolor[rgb]{0,0.8,0.2}-0.00163 & \cellcolor[rgb]{0.212,0.871,0.165}-0.00064 & \cellcolor[rgb]{0.188,0.863,0.169}-0.00067 & \cellcolor[rgb]{0,0.8,0.2}-0.00111 & \cellcolor[rgb]{0,0.8,0.2}-0.00108 &  & \cellcolor[rgb]{0,0.8,0.2}-0.00171^{**} \\ 
				GLASSO-CV2 & \cellcolor[rgb]{1,0.27,0}0.00105 & \cellcolor[rgb]{1,0.27,0}0.00109 & \cellcolor[rgb]{1,0.27,0}0.00088 & \cellcolor[rgb]{1,0.394,0}0.00065 & \cellcolor[rgb]{0.88,0.92,0.24}0.00007 & \cellcolor[rgb]{1,0.27,0}0.00106 & \cellcolor[rgb]{1,0.27,0}0.00104 & \cellcolor[rgb]{1,0.442,0}0.00059 & \cellcolor[rgb]{1,0.412,0}0.00063 & \cellcolor[rgb]{1,0.27,0}0.00171 &  \\ \midrule
				better than \% of models & 0.3 & 0.1 & 0.5 & 0.6 & 0.9 & 0.2 & 0.4 & 0.8 & 0.7 & 0.0 & 1.0 \\ 
				\bottomrule
			\end{tabular}
			\begin{tablenotes}
				\item This table shows the differences in the annualized out-of-sample SD of the GMV-NOSHORT-50SP portfolios across the main covariance estimation methods and their CV2-based counterparts. The table is constructed in a symmetrical way with an applied color scheme from red (higher SD than the other model) to green (lower SD than the other model). In addition, on the elements above the diagonal, the significant pairwise outperformance in terms of variance is denoted by asterisks: *** denotes significance at the 0.001 level; ** denotes significance at the 0.01 level; and * denotes significance at the 0.05 level. Finally, for each model, we report the percentage of the other models that exhibit higher variance as a qualitative measure.
			\end{tablenotes}
		\end{threeparttable}
	}
	\vspace{1cm}
	\caption{Differences in SD p.a. of GMV-NOSHORT-100SP across different estimators. \label{table:sign_100sp_noshort}}
	\resizebox{\linewidth}{!}{%
		\begin{threeparttable}
			\begin{tabular}{l *{11}{d{2.6}}}
				\toprule
				& \mc{Sample} &\mc{LW\textsubscript{1}} & \mc{LW\textsubscript{1}-CV2} & \mc{LW\textsubscript{CC}} & \mc{LW\textsubscript{CC}-CV2} & \mc{LW\textsubscript{NL}} & \mc{LW\textsubscript{NL}-CV2} & \mc{POET} & \mc{POET-CV2} & \mc{GLASSO} & \mc{GLASSO-CV2} \\
				\midrule
				Sample &  & \cellcolor[rgb]{0.987,1,0.003}-0.00007 & \cellcolor[rgb]{1,0.753,0}0.00013 & \cellcolor[rgb]{0.957,1,0.011}-0.00008 & \cellcolor[rgb]{1,0.78,0}0.00009 & \cellcolor[rgb]{0.651,1,0.087}-0.00017 & \cellcolor[rgb]{0.547,0.982,0.109}-0.00019 & \cellcolor[rgb]{0.172,0.857,0.171}-0.00029^{**} & \cellcolor[rgb]{0,0.8,0.2}-0.00044^{***} & \cellcolor[rgb]{0.227,0.876,0.162}-0.00028 & \cellcolor[rgb]{0,0.8,0.2}-0.00104^{**} \\ 
				LW\textsubscript{1} & \cellcolor[rgb]{1,0.8,0}0.00007 &  & \cellcolor[rgb]{1,0.703,0}0.00020 & \cellcolor[rgb]{0.828,1,0.344}-0.00001 & \cellcolor[rgb]{1,0.73,0}0.00016 & \cellcolor[rgb]{0.877,1,0.031}-0.00010 & \cellcolor[rgb]{0.779,1,0.055}-0.00013 & \cellcolor[rgb]{0.427,0.942,0.129}-0.00023 & \cellcolor[rgb]{0,0.8,0.2}-0.00038^* & \cellcolor[rgb]{0.482,0.961,0.12}-0.00021 & \cellcolor[rgb]{0,0.8,0.2}-0.00097^{**} \\ 
				LW\textsubscript{1}-CV2 & \cellcolor[rgb]{0.775,1,0.056}-0.00013 & \cellcolor[rgb]{0.543,0.981,0.11}-0.00020 &  & \cellcolor[rgb]{0.509,0.97,0.115}-0.00020 & \cellcolor[rgb]{0.915,1,0.17}-0.00004 & \cellcolor[rgb]{0.164,0.855,0.173}-0.00030 & \cellcolor[rgb]{0.054,0.818,0.191}-0.00032 & \cellcolor[rgb]{0,0.8,0.2}-0.00042^* & \cellcolor[rgb]{0,0.8,0.2}-0.00057^{**} & \cellcolor[rgb]{0,0.8,0.2}-0.00041 & \cellcolor[rgb]{0,0.8,0.2}-0.00117^{***} \\ 
				LW\textsubscript{CC} & \cellcolor[rgb]{1,0.793,0}0.00008 & \cellcolor[rgb]{0.826,0.974,0.347}0.00001 & \cellcolor[rgb]{1,0.696,0}0.00020 &  & \cellcolor[rgb]{1,0.724,0}0.00017 & \cellcolor[rgb]{0.907,1,0.023}-0.00009 & \cellcolor[rgb]{0.809,1,0.048}-0.00012 & \cellcolor[rgb]{0.46,0.953,0.123}-0.00022 & \cellcolor[rgb]{0,0.8,0.2}-0.00037 & \cellcolor[rgb]{0.516,0.972,0.114}-0.00020 & \cellcolor[rgb]{0,0.8,0.2}-0.00096^{***} \\ 
				LW\textsubscript{CC}-CV2 & \cellcolor[rgb]{0.898,1,0.026}-0.00009 & \cellcolor[rgb]{0.672,1,0.082}-0.00016 & \cellcolor[rgb]{0.909,0.891,0.183}0.00004 & \cellcolor[rgb]{0.642,1,0.089}-0.00017 &  & \cellcolor[rgb]{0.302,0.901,0.15}-0.00026 & \cellcolor[rgb]{0.192,0.864,0.168}-0.00029 & \cellcolor[rgb]{0,0.8,0.2}-0.00039 & \cellcolor[rgb]{0,0.8,0.2}-0.00054 & \cellcolor[rgb]{0,0.8,0.2}-0.00037 & \cellcolor[rgb]{0,0.8,0.2}-0.00113^{***} \\ 
				LW\textsubscript{NL} & \cellcolor[rgb]{1,0.725,0}0.00017 & \cellcolor[rgb]{1,0.776,0}0.00010 & \cellcolor[rgb]{1,0.628,0}0.00030 & \cellcolor[rgb]{1,0.782,0}0.00009 & \cellcolor[rgb]{1,0.656,0}0.00026 &  & \cellcolor[rgb]{0.892,1,0.217}-0.00003 & \cellcolor[rgb]{0.783,1,0.054}-0.00013 & \cellcolor[rgb]{0.235,0.878,0.161}-0.00028 & \cellcolor[rgb]{0.832,1,0.042}-0.00011 & \cellcolor[rgb]{0,0.8,0.2}-0.00087^{**} \\ 
				LW\textsubscript{NL}-CV2 & \cellcolor[rgb]{1,0.704,0}0.00019 & \cellcolor[rgb]{1,0.754,0}0.00013 & \cellcolor[rgb]{1,0.607,0}0.00032 & \cellcolor[rgb]{1,0.761,0}0.00012 & \cellcolor[rgb]{1,0.634,0}0.00029 & \cellcolor[rgb]{0.886,0.914,0.227}0.00003 &  & \cellcolor[rgb]{0.88,1,0.03}-0.00010 & \cellcolor[rgb]{0.345,0.915,0.143}-0.00025 & \cellcolor[rgb]{0.93,1,0.018}-0.00008 & \cellcolor[rgb]{0,0.8,0.2}-0.00084^{**} \\ 
				POET & \cellcolor[rgb]{1,0.63,0}0.00029 & \cellcolor[rgb]{1,0.68,0}0.00023 & \cellcolor[rgb]{1,0.533,0}0.00042 & \cellcolor[rgb]{1,0.687,0}0.00022 & \cellcolor[rgb]{1,0.56,0}0.00039 & \cellcolor[rgb]{1,0.755,0}0.00013 & \cellcolor[rgb]{1,0.776,0}0.00010 &  & \cellcolor[rgb]{0.707,1,0.073}-0.00015^* & \cellcolor[rgb]{0.844,0.956,0.313}0.00001 & \cellcolor[rgb]{0,0.8,0.2}-0.00075^{**} \\ 
				POET-CV2 & \cellcolor[rgb]{1,0.518,0}0.00044 & \cellcolor[rgb]{1,0.568,0}0.00038 & \cellcolor[rgb]{1,0.42,0}0.00057 & \cellcolor[rgb]{1,0.574,0}0.00037 & \cellcolor[rgb]{1,0.448,0}0.00054 & \cellcolor[rgb]{1,0.642,0}0.00028 & \cellcolor[rgb]{1,0.664,0}0.00025 & \cellcolor[rgb]{1,0.738,0}0.00015 &  & \cellcolor[rgb]{1,0.727,0}0.00016 & \cellcolor[rgb]{0,0.8,0.2}-0.00060^* \\ 
				GLASSO & \cellcolor[rgb]{1,0.641,0}0.00028 & \cellcolor[rgb]{1,0.691,0}0.00021 & \cellcolor[rgb]{1,0.544,0}0.00041 & \cellcolor[rgb]{1,0.698,0}0.00020 & \cellcolor[rgb]{1,0.571,0}0.00037 & \cellcolor[rgb]{1,0.766,0}0.00011 & \cellcolor[rgb]{1,0.787,0}0.00008 & \cellcolor[rgb]{0.846,1,0.307}-0.00001 & \cellcolor[rgb]{0.657,1,0.086}-0.00016 &  & \cellcolor[rgb]{0,0.8,0.2}-0.00076^{**} \\ 
				GLASSO-CV2 & \cellcolor[rgb]{1,0.27,0}0.00104 & \cellcolor[rgb]{1,0.27,0}0.00097 & \cellcolor[rgb]{1,0.27,0}0.00117 & \cellcolor[rgb]{1,0.27,0}0.00096 & \cellcolor[rgb]{1,0.27,0}0.00113 & \cellcolor[rgb]{1,0.27,0}0.00087 & \cellcolor[rgb]{1,0.27,0}0.00084 & \cellcolor[rgb]{1,0.29,0}0.00075 & \cellcolor[rgb]{1,0.402,0}0.00060 & \cellcolor[rgb]{1,0.279,0}0.00076 &  \\ \midrule
				better than \% of models & 0.2 & 0.3 & 0.0 & 0.4 & 0.1 & 0.5 & 0.6 & 0.8 & 0.9 & 0.7 & 1.0 \\ 
				\bottomrule
			\end{tabular}
			\begin{tablenotes}
				\item This table shows the differences in the annualized out-of-sample SD of the GMV-NOSHORT-100SP portfolios across the main covariance estimation methods and their CV2-based counterparts. The table is constructed in a symmetrical way with an applied color scheme from red (higher SD than the other model) to green (lower SD than the other model). In addition, on the elements above the diagonal, the significant pairwise outperformance in terms of variance is denoted by asterisks: *** denotes significance at the 0.001 level; ** denotes significance at the 0.01 level; and * denotes significance at the 0.05 level. Finally, for each model, we report the percentage of the other models that exhibit higher variance as a qualitative measure.    
			\end{tablenotes}
		\end{threeparttable}
	}
\end{sidewaystable}

\begin{sidewaystable}[pt!]
	\centering
	\caption{Differences in SD p.a. of GMV-NOSHORT-200SP across different estimators.\label{table:sign_200sp_noshort}}
	\resizebox{\linewidth}{!}{%
		\begin{threeparttable}
			\begin{tabular}{l *{11}{d{2.6}}}
				\toprule
				& \mc{Sample} &\mc{LW\textsubscript{1}} & \mc{LW\textsubscript{1}-CV2} & \mc{LW\textsubscript{CC}} & \mc{LW\textsubscript{CC}-CV2} & \mc{LW\textsubscript{NL}} & \mc{LW\textsubscript{NL}-CV2} & \mc{POET} & \mc{POET-CV2} & \mc{GLASSO} & \mc{GLASSO-CV2} \\
				\midrule
				Sample &  & \cellcolor[rgb]{0,0.8,0.2}-0.00027 & \cellcolor[rgb]{0.273,0.891,0.154}-0.00018 & \cellcolor[rgb]{0,0.8,0.2}-0.00037 & \cellcolor[rgb]{0,0.8,0.2}-0.00030 & \cellcolor[rgb]{0.299,0.9,0.15}-0.00017 & \cellcolor[rgb]{0.139,0.846,0.177}-0.00022 & \cellcolor[rgb]{0.003,0.801,0.2}-0.00026 & \cellcolor[rgb]{0,0.8,0.2}-0.00038^{***} & \cellcolor[rgb]{0,0.8,0.2}-0.00031 & \cellcolor[rgb]{0,0.8,0.2}-0.00056 \\ 
				LW\textsubscript{1} & \cellcolor[rgb]{1,0.278,0}0.00027 &  & \cellcolor[rgb]{1,0.734,0}0.00009 & \cellcolor[rgb]{0.522,0.974,0.113}-0.00010 & \cellcolor[rgb]{0.718,1,0.276}-0.00003 & \cellcolor[rgb]{1,0.713,0}0.00010 & \cellcolor[rgb]{1,0.845,0}0.00005 & \cellcolor[rgb]{1,0.968,0}0.00001 & \cellcolor[rgb]{0.489,0.963,0.119}-0.00011 & \cellcolor[rgb]{0.687,1,0.231}-0.00004 & \cellcolor[rgb]{0,0.8,0.2}-0.00029 \\ 
				LW\textsubscript{1}-CV2 & \cellcolor[rgb]{1,0.523,0}0.00018 & \cellcolor[rgb]{0.541,0.98,0.11}-0.00009 &  & \cellcolor[rgb]{0.213,0.871,0.165}-0.00019 & \cellcolor[rgb]{0.437,0.946,0.127}-0.00013 & \cellcolor[rgb]{1,0.98,0}0.00001 & \cellcolor[rgb]{0.693,1,0.239}-0.00004 & \cellcolor[rgb]{0.579,0.993,0.103}-0.00008 & \cellcolor[rgb]{0.179,0.86,0.17}-0.00020 & \cellcolor[rgb]{0.4,0.933,0.133}-0.00014 & \cellcolor[rgb]{0,0.8,0.2}-0.00039 \\ 
				LW\textsubscript{CC} & \cellcolor[rgb]{1,0.27,0}0.00037 & \cellcolor[rgb]{1,0.719,0}0.00010 & \cellcolor[rgb]{1,0.475,0}0.00019 &  & \cellcolor[rgb]{1,0.801,0}0.00007 & \cellcolor[rgb]{1,0.454,0}0.00020 & \cellcolor[rgb]{1,0.58,0}0.00015 & \cellcolor[rgb]{1,0.689,0}0.00011 & \cellcolor[rgb]{0.773,1,0.36}-0.00001 & \cellcolor[rgb]{1,0.835,0}0.00006 & \cellcolor[rgb]{0.213,0.871,0.165}-0.00019 \\ 
				LW\textsubscript{CC}-CV2 & \cellcolor[rgb]{1,0.27,0}0.00030 & \cellcolor[rgb]{1,0.91,0}0.00003 & \cellcolor[rgb]{1,0.652,0}0.00013 & \cellcolor[rgb]{0.62,1,0.13}-0.00007 &  & \cellcolor[rgb]{1,0.632,0}0.00013 & \cellcolor[rgb]{1,0.758,0}0.00008 & \cellcolor[rgb]{1,0.875,0}0.00004 & \cellcolor[rgb]{0.592,0.997,0.101}-0.00008 & \cellcolor[rgb]{0.77,1,0.355}-0.00001 & \cellcolor[rgb]{0,0.8,0.2}-0.00026 \\ 
				LW\textsubscript{NL} & \cellcolor[rgb]{1,0.543,0}0.00017 & \cellcolor[rgb]{0.515,0.972,0.114}-0.00010 & \cellcolor[rgb]{0.779,1,0.369}-0.00001 & \cellcolor[rgb]{0.187,0.862,0.169}-0.00020 & \cellcolor[rgb]{0.412,0.937,0.131}-0.00013 &  & \cellcolor[rgb]{0.672,1,0.208}-0.00005^{**} & \cellcolor[rgb]{0.553,0.984,0.108}-0.00009 & \cellcolor[rgb]{0.153,0.851,0.174}-0.00021 & \cellcolor[rgb]{0.374,0.925,0.138}-0.00014 & \cellcolor[rgb]{0,0.8,0.2}-0.00040 \\ 
				LW\textsubscript{NL}-CV2 & \cellcolor[rgb]{1,0.417,0}0.00022 & \cellcolor[rgb]{0.66,1,0.189}-0.00005 & \cellcolor[rgb]{1,0.883,0}0.00004 & \cellcolor[rgb]{0.347,0.916,0.142}-0.00015 & \cellcolor[rgb]{0.571,0.99,0.105}-0.00008 & \cellcolor[rgb]{1,0.859,0}0.00005 &  & \cellcolor[rgb]{0.691,1,0.236}-0.00004 & \cellcolor[rgb]{0.313,0.904,0.148}-0.00016 & \cellcolor[rgb]{0.534,0.978,0.111}-0.00010 & \cellcolor[rgb]{0,0.8,0.2}-0.00035 \\ 
				POET & \cellcolor[rgb]{1,0.309,0}0.00026 & \cellcolor[rgb]{0.769,1,0.354}-0.00001 & \cellcolor[rgb]{1,0.764,0}0.00008 & \cellcolor[rgb]{0.483,0.961,0.119}-0.00011 & \cellcolor[rgb]{0.687,1,0.23}-0.00004 & \cellcolor[rgb]{1,0.744,0}0.00009 & \cellcolor[rgb]{1,0.88,0}0.00004 &  & \cellcolor[rgb]{0.45,0.95,0.125}-0.00012 & \cellcolor[rgb]{0.656,1,0.185}-0.00005 & \cellcolor[rgb]{0,0.8,0.2}-0.00031 \\ 
				POET-CV2 & \cellcolor[rgb]{1,0.27,0}0.00038 & \cellcolor[rgb]{1,0.693,0}0.00011 & \cellcolor[rgb]{1,0.448,0}0.00020 & \cellcolor[rgb]{1,0.973,0}0.00001 & \cellcolor[rgb]{1,0.774,0}0.00008 & \cellcolor[rgb]{1,0.428,0}0.00021 & \cellcolor[rgb]{1,0.554,0}0.00016 & \cellcolor[rgb]{1,0.662,0}0.00012 &  & \cellcolor[rgb]{1,0.805,0}0.00007 & \cellcolor[rgb]{0.246,0.882,0.159}-0.00018 \\ 
				GLASSO & \cellcolor[rgb]{1,0.27,0}0.00031 & \cellcolor[rgb]{1,0.876,0}0.00004 & \cellcolor[rgb]{1,0.622,0}0.00014 & \cellcolor[rgb]{0.65,1,0.176}-0.00006 & \cellcolor[rgb]{1,0.969,0}0.00001 & \cellcolor[rgb]{1,0.602,0}0.00014 & \cellcolor[rgb]{1,0.728,0}0.00010 & \cellcolor[rgb]{1,0.841,0}0.00005 & \cellcolor[rgb]{0.624,1,0.136}-0.00007 &  & \cellcolor[rgb]{0.025,0.808,0.196}-0.00025 \\ 
				GLASSO-CV2 & \cellcolor[rgb]{1,0.27,0}0.00056 & \cellcolor[rgb]{1,0.27,0}0.00029 & \cellcolor[rgb]{1,0.27,0}0.00039 & \cellcolor[rgb]{1,0.475,0}0.00019 & \cellcolor[rgb]{1,0.297,0}0.00026 & \cellcolor[rgb]{1,0.27,0}0.00040 & \cellcolor[rgb]{1,0.27,0}0.00035 & \cellcolor[rgb]{1,0.27,0}0.00031 & \cellcolor[rgb]{1,0.501,0}0.00018 & \cellcolor[rgb]{1,0.327,0}0.00025 &  \\ \midrule
				better than \% of models & 0.0 & 0.5 & 0.2 & 0.8 & 0.6 & 0.1 & 0.3 & 0.4 & 0.9 & 0.7 & 1.0 \\ 
				\bottomrule
			\end{tabular}
			\begin{tablenotes}
				\item This table shows the differences in the annualized out-of-sample SD of the GMV-NOSHORT-200SP portfolios across the main covariance estimation methods and their CV2-based counterparts. The table is constructed in a symmetrical way with an applied color scheme from red (higher SD than the other model) to green (lower SD than the other model). In addition, on the elements above the diagonal, the significant pairwise outperformance in terms of variance is denoted by asterisks: *** denotes significance at the 0.001 level; ** denotes significance at the 0.01 level; and * denotes significance at the 0.05 level. Finally, for each model, we report the percentage of the other models that exhibit higher variance as a qualitative measure.
			\end{tablenotes}
		\end{threeparttable}
	}
\end{sidewaystable}

\newpage
\section*{Short Biographies}

\noindent Sven Husmann 

\noindent Sven Husmann is Professor of Finance at the European University Viadrina in Frankfurt (Oder), Germany. He has been a faculty member since 2002 and was Vice President of the University from 2014-2018. He received his doctorate at the FU Berlin in 2001 with a thesis on option pricing and held an assistant professorship in International Accounting from 2002-2007. His research interests include option pricing, company valuation and portfolio management.

\vspace{1cm}

\noindent Antoniya Shivarova

\noindent Antoniya Shivarova works as a research assistant at the chair of Finance and Capital Market Theory of the European University Viadrina in Frankfurt (Oder), Germany, where she received a M.Sc. and B.Sc. degree in Business Administration with focus on finance. Apart from her activities as a research assistant, Antoniya Shivarova takes part in the Ph.D. program of the European University Viadrina. Her research focus lies on portfolio management and covariance estimation. 

\vspace{1cm}

\noindent Rick Steinert

\noindent Rick Steinert works as a research assistant at the chair of finance and capital market theory of
the European University Viadrina in Frankfurt (Oder), Germany, where he received a M.Sc. and B.Sc. degree in business administration with focus on finance. Apart from his activities as a research assistant, Rick Steinert takes part in the Ph.D. program of European University Viadrina. His research focus lies on portfolio management and electricity pricing. 

\end{document}